\documentclass[aps,rmp,amsmath,longbibliography,amssymb,twocolumn]{revtex4-1}
\usepackage{graphicx}
\usepackage{xcolor}
\usepackage{subfigure}
\usepackage{adjustbox}
\usepackage{bm}
\usepackage{color}
\usepackage{braket}
\usepackage{standalone}
\usepackage{multirow}
\usepackage{tikz}
\usepackage{mathrsfs}
\usepackage{dsfont}
\usepackage{bbold}
\usepackage[colorlinks,bookmarks=true,citecolor=blue,linkcolor=red,urlcolor=blue]{hyperref}

\newcommand{\bea}{\begin{eqnarray}}
\newcommand{\eea}{\end{eqnarray}}

\begin{document}

\title{Exceptional Topology of Non-Hermitian Systems}

\author{Emil J. Bergholtz$^{1}$}\email{emil.bergholtz@fysik.su.se}
\author{Jan Carl Budich$^{2}$}
\email{jan.budich@tu-dresden.de}
 \author{Flore K. Kunst$^{1,3}$}
\email{flore.kunst@mpq.mpg.de}
\thanks{\\ Authors in alphabetic order.}
\affiliation{$^1$Department of Physics${\rm ,}$ Stockholm University, AlbaNova University Center${\rm ,}$ 106 91 Stockholm${\rm ,}$ Sweden\\
$^2$Institute of Theoretical Physics${\rm ,}$ Technische Universit\"{a}t Dresden and W\"{u}rzburg-Dresden Cluster of Excellence ct.qmat${\rm ,}$ 01062 Dresden${\rm ,}$ Germany\\
$^3$Max-Planck-Institut f\"{u}r Quantenoptik, Hans-Kopfermann-Stra{\ss}e 1${\rm ,}$ 85748 Garching${\rm ,}$ Germany
}

\date{\today}

\begin{abstract}
The current understanding of the role of topology in non-Hermitian (NH) systems and its far-reaching physical consequences observable in a range of dissipative settings are reviewed. In particular, how the paramount and genuinely NH concept of exceptional degeneracies, at which both eigenvalues and eigenvectors coalesce, leads to phenomena drastically distinct from the familiar Hermitian realm is discussed. An immediate consequence is the ubiquitous occurrence of nodal NH topological phases with concomitant open Fermi-Seifert surfaces, where conventional band-touching points are replaced by the aforementioned exceptional degeneracies. Furthermore, new notions of gapped phases including topological phases in single-band systems are detailed, and the manner in which a given physical context may affect the symmetry-based topological classification is clarified. A unique property of NH systems with relevance beyond the field of topological phases consists of the anomalous relation between bulk and boundary physics, stemming from the striking sensitivity of NH matrices to boundary conditions. Unifying several complementary insights recently reported in this context, a picture of intriguing phenomena such as the NH bulk-boundary correspondence and the NH skin effect is put together. Finally, applications of NH topology in both classical systems including optical setups with gain and loss, electric circuits,s and mechanical systems and genuine quantum systems such as electronic transport settings at material junctions and dissipative cold-atom setups are reviewed.\end{abstract}

\maketitle

{
\hypersetup{linkcolor=black}
  \tableofcontents
}
\newpage

\section{Introduction}
\label{sec:intro}
One of the basic axioms of quantum mechanics requires observables, such as the Hamiltonian of a closed system, to be self-adjoint operators, which are typically represented by Hermitian matrices. Real physical systems, however, are at least to some extent coupled to their environment, where the presence of dissipative processes renders their description more complex: In general, the familiar Schr\"odinger equation with a Hermitian Hamiltonian there is replaced by a Liouvillian superoperator governing the time evolution of the density matrix \cite{Breuer2002}. In certain regimes, such open systems in contact with an environment can be accurately described by approaches such as Lindblad quantum master equations \cite{Lindblad1976}, Feynman-Vernon theory \cite{FEYNMAN1963118}, and the Keldysh formalism \cite{Keldysh:1964ud}. While immensely powerful, the technical complexity of these methods severely limits the range of systems that can be efficiently studied. Effective non-Hermitian (NH) Hamiltonians provide a conceptually simpler and intuitive alternative to fully microscopic approaches, and have already led to profound insights with applications. The spectrum of physical platforms ranges from classical systems, including optical settings, electrical circuits, and mechanical systems, which may be mapped to an effective NH Schr\"odinger equation, all the way to quantum materials \cite{Miri2019,Rotter2009,El-Ganainy2018,Bender2007,Ozawa2019,Datta2005}.

In a wider historical context, effective NH concepts have been ubiquitous for many decades \cite{Majorana1931,Pancharatnam1955,KatoBook,Hatano1996, Berry1998,Berry2004,Efetov,PietPeter,Efetov2,HatanoNelson2,Silvestrov1998,Silvestrov1999} for example for describing resonances and broadening in scattering problems in atomic and particle physics, as well as in nuclear reactions \cite{Majorana1931,Majorana1931b,Fano,BreitWigner,FESHBACH1958357,FeshbachRev}, all the way to applications in biological systems \cite{NHBIO,NHDNA}. Following the seminal insight that NH Hamiltonians preserving the combination of parity and time-reversal ($PT$) symmetry stably feature real spectra \cite{Bender1998,Bender2007}, relinquishing the assumption of Hermiticity has even been considered a fundamental amendment to quantum physics. By now, $PT$-symmetric Hamiltonians are well established as an effective description of dissipative systems with balanced gain and loss \cite{El-Ganainy2018}.

In parallel to these developments, the advent of topological phases such as topological insulators and semimetals has revolutionized the classification of matter and led to groundbreaking discoveries of topologically robust physical phenomena \cite{HasanKane,QiZhang,ChiuRev,Armitage2018}. With motivation provided by experiments reporting novel topological states in dissipative settings \cite{Zhou2018,Bandres2018,NHexp,Cerjan2019,Poli2015Selective,asymhop2, Hodaei2017,Zeuner2015, HeHoImAbKiMoLeSzGrTh2019}, extending the notion of topological phases to NH systems has become a broad frontier of current research. In this context a plethora of uniquely non-Hermitian aspects of topological systems have been revealed \cite{gong,KuEdBuBe2018,Kawabata2018,YaWa2018}. Salient examples in the focus of our review include an anomalous bulk-boundary correspondence accompanied by the non-Hermitian skin effect \cite{KuEdBuBe2018,YaWa2018,Martinez2018,Xiong2018,LeeSkinEffect2016}, the ubiquitous occurrence of exceptional nodal phases \cite{koziifu,Budich2019,Okugawa2019,Zhou2018,Yoshida2019a,Szameit2011,Rui2019} with open Fermi-Seifert surfaces \cite{knots,carlstroembergholtz,knots2}, and a new system of generic symmetries \cite{Bernard2002} forming the basis for the topological classification of both gapless \cite{Budich2019,Kawabata2019} and gapped \cite{Esaki2011,Kawabata2018,LieuSym,ZhouPeriodicTable, ShenFu, Leykam} NH topological phases. In this review, we provide a comprehensive overview of these developments with an emphasis on their relation to exceptional degeneracies at which both eigenvalues and eigenvectors coalesce, a paramount spectral feature unique to NH systems.

\emph{Exceptional degeneracies in NH two-level systems.}---As preparation for the NH Hamiltonian formalism to be detailed, we discuss a minimal two-level example that may serve as an intuitive basis for understanding many of the key concepts unique to NH matrices, in particular, the aforementioned exceptional degeneracies. Specifically, we consider the effective Hamiltonian
\begin{equation}
H = \begin{pmatrix}
0 & \alpha \\
1 & 0
\end{pmatrix}, \qquad \alpha \in \mathbb{C},
\end{equation}
whose complex energy eigenvalues 
\begin{equation}
E_\pm = \pm \sqrt{\alpha}
\label{eqn:toyevals}
\end{equation}
generate a generically nonunitary time evolution. Another key observation is that the eigenspectra of NH systems are not analytic in the system parameters due to the divergence of $|\partial_\alpha E(\alpha)|\rightarrow \infty$ as $\alpha\rightarrow 0$, which has been proposed as a mechanism for new sensing devices \cite{asymhop2, Hodaei2017}.
In contrast to the Hermitian case, the right eigenvectors defined by $H \psi_{R,\pm} = E_{\pm}  \psi_{R,\pm}$ and left eigenvectors satisfying $\psi_{L,\pm} H  = \psi_{L,\pm} E_{\pm}$ are generically different. Here explicitly
\begin{equation}
\psi_{R, \pm} = \begin{pmatrix}
\pm\sqrt{\alpha} \\
 1
\end{pmatrix}, \qquad \psi_{L, \pm} = \begin{pmatrix}
1 & \pm \sqrt{\alpha}
\end{pmatrix}.
\label{eqn:toyevecs}
\end{equation}
Hence, in clear contrast to Hermitian Hamiltonians, $\psi_{R, \pm}\neq \psi_{L, \pm}^\dagger$ while $\psi_{R/L,+}$ and $\psi_{R/L,-}$ are not orthogonal for $\alpha\neq 1$.
At the {\it exceptional point} (EP), $\alpha = 0$, $H$ assumes a Jordan block form, and, in addition to the twofold degeneracy of the eigenvalue at $E=0$, the eigenvectors coalesce such that only a single right and a single left eigenvector remain \cite{Heiss2012}; see Eq. (\ref{eqn:toyevecs}). On a more technical note, at the EP the matrix $H$ becomes defective, meaning that the geometric multiplicity (number of linearly independent eigenvectors) is smaller than the algebraic multiplicity (degree of degeneracy in the characteristic polynomial) for the eigenvalue $E=0$. 

To better understand the consequences of this scenario, we consider tracing a loop in the complex plane with the parameter $\alpha$, so as to enclose the EP at $\alpha=0$. With $\alpha=|\alpha|e^{i\arg(\alpha)}$ we have $E=\pm|\alpha|^{1/2}e^{i\arg(\alpha)/2}$ with $-\pi <\arg(\alpha)\leq\pi$ on the principal domain. Note that away from the EP there is always a finite complex-energy gap $\Delta E=2|\alpha|^{1/2}e^{i\arg(\alpha)/2}$, and one can thus unambiguously track the energies and the corresponding eigenstates. However, following an eigenstate and its corresponding energy while encircling the exceptional point through $\arg(\alpha)\rightarrow \arg(\alpha)+2\pi$ one readily finds that
 \begin{equation}\psi_{R/L, \pm}\rightarrow \psi_{R/L, \mp}, \qquad E_ \pm\rightarrow E_{\mp}\ . \end{equation} 
This swapping of eigenvalues, as a manifestation of the complex energy living on a two-sheeted Riemann surface known from the behavior of the complex square-root function around the origin, is directly associated with the presence of second-order exceptional points; cf. Eq.~(\ref{eqn:toyevals}). A striking implication is that while encircling the EP at $\alpha=0$, the real part of the energy crosses zero exactly once, namely when it passes the branch cut on the negative real line, i.e., at $\arg(\alpha)=\pi$. In Sec.~\ref{sec:NH_band_theory}, precisely this property is shown to lead to the occurrence of novel NH Fermi arcs, and higher-dimensional generalizations thereof, as a unique and ubiquitous feature of NH band structures. 

The remainder of this review article is organized as follows. In Sec.~\ref{sec:NH_band_theory}, we discuss in detail the topological band theory of non-Hermitian systems including both nodal phases, which are found to be much more abundant than in the Hermitian realm, and various notions of gapped systems generalizing the concept of insulators. In Sec.~\ref{sec:NH_BBC}, we discuss how the bulk-boundary correspondence, i.e., the direct relation between bulk topological invariants and the occurrence of protected surface states, is qualitatively modified in NH systems. This phenomenon is shown to be closely related to the NH skin effect, i.e., the accumulation of a macroscopic number of eigenstates at the boundary of systems with open boundary conditions. In both Sec.~\ref{sec:NH_band_theory} and Sec.~\ref{sec:NH_BBC}, we clarify the direct relation of the uniquely NH phenomenology to the presence or proximity of EPs. In Sec.~\ref{sec:NH_applications}, we then give an overview of both classical and quantum systems in which the fundamental aspects of NH topology have been predicted to occur or have even already been experimentally demonstrated. A concluding discussion is presented in Sec.~\ref{sec:conclusion}, providing an outlook toward a conclusive understanding of the role of topology in NH systems.

Throughout this review we aim for a self-contained presentation; however, a basic knowledge of Hermitian topological band structures is helpful, for which we refer the interested reader to reviews 
\cite{HasanKane,QiZhang,ChiuRev,Armitage2018}.

\section{Non-Hermitian topological band theory}
\label{sec:NH_band_theory}
In this section, we systematically review the topological properties of Bloch bands in NH systems. The recent pursuit of topologically classifying NH band structures has led to the experimental discovery and theoretical explanation of various topologically stable phenomena that have no direct counterpart in the Hermitian realm, including a novel system of gapped and gapless (symmetry-protected) NH topological phases discussed in this section. 

\subsection{Basic concepts and minimal examples}
\label{sec:NHBandsBasics}
To get an intuitive feeling for the topological properties of NH Bloch bands, we start by discussing some elementary examples.

\begin{figure*}
  \includegraphics[width=0.9\textwidth]{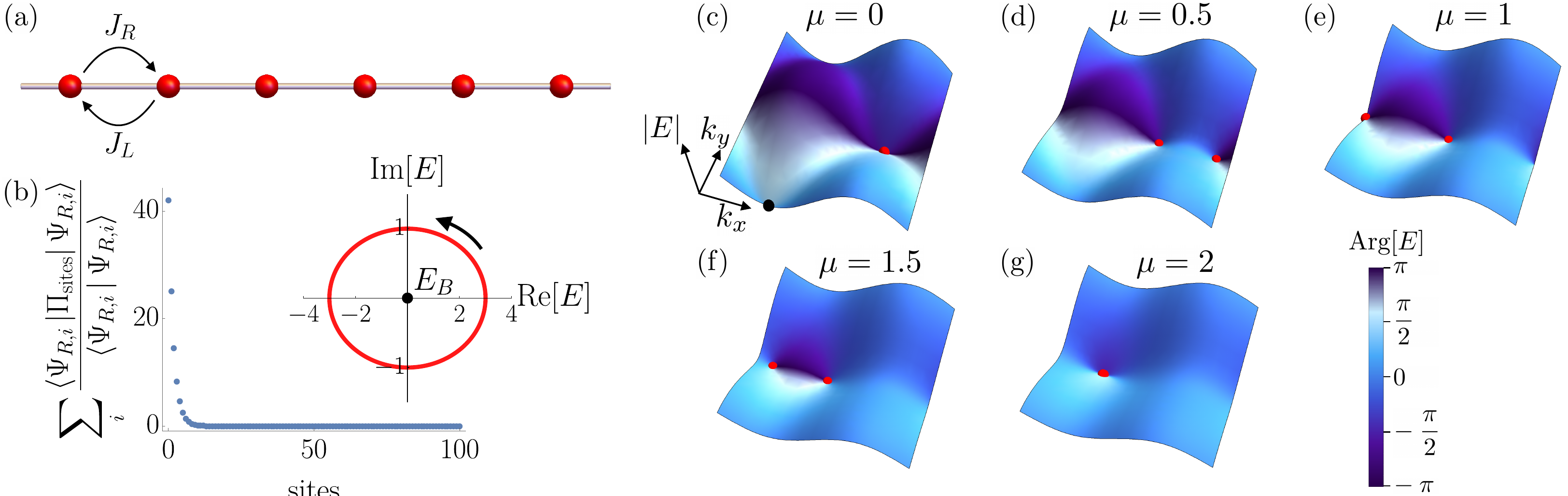}
  \caption{(a) Schematic depiction of the Hatano-Nelson model; see Eq.~(\ref{eq:hatano_nelson_model}). (b) Sum of absolute squares of amplitudes per site of all right eigenstates for the Hamiltonian in Eq.~(\ref{eq:hatano_nelson_model}) with OBCs for $100$ sites and $|J_L|/|J_R| = 2$. Inset: energy in the complex plane, which winds around the base point $E_B$ in the complex plane with winding number $w=1$ when $|J_L|>|J_R|$ (in this case $|J_L|/|J_R| = 2$). (c)--(g) Absolute value of Eq.~(\ref{Emu}) for different values of $\mu$, as indicated: (c) $\mu = 0$. (d) $\mu = 0.5$. (e) $\mu = 1$. (f) $\mu = 1.5$. (g) $\mu = 2$. The plotting axes are shown in (c), and the color corresponds to the argument of $E_\mu$ with the color bar in (g). For $\mu$ increasing from zero, two vortices ($E_\mu = 0$) split and merge again when $\mu \rightarrow 2$. The vortices are shown by red (gray) spheres. There is an additional zero-energy solution when $\mu=0$, which is shown with a black sphere in (c), at $(k_x,k_y)=(-\pi/2,0)$, which is gapped rather than split upon increasing $\mu$. Note that if we were to decrease $\mu$ from $0$ to $-2$, it would be this (black) zero that splits into vortices, and the other zero (red) would be gapped out.}
  \label{fig:hatano_nelson_model}
\end{figure*}

\subsubsection{Topological one-band models}
\label{sec:one_band_NH}
\textit{Hatano-Nelson model.}---In sharp contrast to Hermitian systems, even a band structure consisting only of a {\emph{single}} band may be topologically nontrivial in the NH context. A paradigmatic example along these lines is provided by the Hatano-Nelson model, which was initially proposed to study localization transitions in superconductors \cite{Hatano1996},
\begin{equation}
H= \sum_{n} \left(J_Lc^\dagger_nc_{n+1}+J_Rc^\dagger_{n+1}c_{n}\right), \qquad J_L, J_R \in \mathbb{R}, \label{eq:hatano_nelson_model}
\end{equation}
where $c^\dagger_n$ ($c_n$) creates (annihilates) a state on site $n$, and with $|J_L|\neq |J_R|$ in general; see Fig.~\ref{fig:hatano_nelson_model}(a). The complex-energy spectrum reads as $E_k = \left(J_L + J_R\right) \textrm{cos}(k) + i \left(J_L - J_R\right) \textrm{sin}(k)$ and, as a function of $k$, winds around the origin in the clockwise (counterclockwise) direction when $|J_L |- |J_R|<0$ ($|J_L |- |J_R|>0$), as shown in the inset of Fig.~\ref{fig:hatano_nelson_model}(b). These phases are formally (homotopically) distinguished by the integer quantized value $w = -1$ ($1$) of the spectral winding number \cite{ShenFu,gong}
\begin{equation}
w = \frac{1}{2 \pi i} \int_{-\pi}^\pi d k \, \partial_k \, \textrm{ln} \, E_k.
\label{eqn:spectralwinding}
\end{equation}

A transition between the two topologically distinct regimes then requires $E_k=0$ for some $k$ (here at $|J_L| = |J_R|$). We stress the conceptual difference between the topological invariant (\ref{eqn:spectralwinding}), which distinguishes inequivalent paths in the complex-\emph{energy} plane, and standard Hermitian topological invariants, which quantify some winding of the {\textit{eigenstates}} based on the Berry connection. On a more technical note, the Hatano-Nelson model (\ref{eq:hatano_nelson_model}) represents a minimal example of a system with a so-called point gap around the singular point $E=0$ in the spectrum \cite{Kawabata2018}; see Sec.~\ref{sec:gapped} for a more detailed discussion. For general multiband models, we note that $E_k$ is simply replaced by $\text{det} H(k)$ in Eq.~(\ref{eqn:spectralwinding}), where $H(k)$ denotes the effective NH Hamiltonian in reciprocal space (Bloch Hamiltonian), such that the winding number in Eq.~(\ref{eqn:spectralwinding}) generically has integer ($\mathbb{Z}$) values.

{\textit{Non-Hermitian skin effect.}}---The asymmetric hopping strength $|J_R|\neq |J_L|$ in the Hatano-Nelson model [Eq.~(\ref{eq:hatano_nelson_model})] gives rise to another exotic feature unique to NH systems: In the case of open boundary conditions, a macroscopic number of eigenstates pile up at one of the ends, a phenomenon known as the non-Hermitian skin effect \cite{Xiong2018, YaWa2018, MartinezAlvarez2018, KuEdBuBe2018}; cf. Fig.~\ref{fig:hatano_nelson_model}(b). The end at which the weight of the eigenstates accumulates depends on which direction of hopping is dominant. This becomes particularly intuitive when one of the hopping directions is entirely turned off, e.g., $J_L= 0$ in Eq.~(\ref{eq:hatano_nelson_model}). In this case the Hamiltonian with open boundary conditions can be written as a single Jordan block such that the energy spectrum features an exceptional point of order $N$, where $N$ is the total number of sites. The proximity to such high-order exceptional points, the order of which scales with system size \cite{Martinez2018,Kunst2019}, in generic models with open boundary conditions are at the heart of the breakdown of the conventional bulk-boundary correspondence as discussed in Sec.~\ref{sec:NH_BBC}.

{\textit{Complex-energy vortices.}}---It is natural to consider higher-dimensional extensions of the Hatano-Nelson model. There we find that zeros in the spectrum lead to the formation of topologically stable vortices in the complex energy. For instance, consider the single-band non-Hermitian nearest-neighbor single-band model corresponding to the spectrum
\begin{equation}
E(\mathbf k)=\sin(k_x)+i\sin(k_y), \label{E}
\end{equation}
which has vortices (zeros) when both momenta are at $0$ or $\pi$ yielding a total of four zeros in the BZ. Focusing on the zero at $\mathbf k=0$ it is clear that it is associated with a finite winding number $
w = 1/(2 \pi i) \oint_C d k \, \partial_k \, \textrm{ln} \, E_k $,
where the closed path $C$ now encloses the origin but no other zeros. This winding has the intriguing consequence that it implies the existence of robust lines of zero real and imaginary energy, connecting the zeros in the spectrum.

A model with a minimal number of two complex zeros can be constructed with 
\begin{equation}
E_\mu(\mathbf k)=\sin(k_x)+\cos(k_y)+\mu+i\sin(k_y), \label{Emu}
\end{equation}
which displays the stability of the vortices upon varying $\mu$: The vortices split at a singular point when $\mu$ is increased from zero, and, after traveling in opposite directions through the BZ, merge again at $\mu\rightarrow 2$, as shown in Figs.~\ref{fig:hatano_nelson_model}(c)-(g).

On a more conceptual note, the considered two-dimensional systems represent a dimensional extension to a gapless topological phase from a point-gapped lower-dimensional model (the Hatano-Nelson model). This phenomenology bears similarities to Weyl semimetals in the Hermitian realm in three spatial dimensions that may be seen as families of Chern insulators in two spatial dimensions, where the Weyl points correspond to topological quantum phase transitions between different Chern numbers \cite{Armitage2018}. To see this analogy, we can rewrite Eq.~(\ref{Emu}) as
\begin{equation}
E_\mu(\mathbf k)=[\sin(k_x)+\mu]+e^{ik_y}, \label{eq:two_dim_gen_hatano_nelson}
\end{equation}
which, seen as a one-dimensional Hatano-Nelson type model at fixed $k_x$, changes its winding number [see Eq.~(\ref{eqn:spectralwinding})] precisely at the position of the complex zeros (vortices). Since the instantaneous 1D model corresponds to unidirectional hopping in the positive $y$ direction and $\sin(k_x)+\mu$ is a simple shift of all energy levels, all eigenstates coalesce and are located at the end site in an open chain geometry. Hence, the aforementioned NH skin effect occurs at all $k_x$, while the winding number also changes as a function of $k_x$, thus highlighting the fact that there is no direct correspondence between these two phenomena unless further assumptions are included, as discussed in Sec.~\ref{subsubsec:skin_effect}.

\subsubsection{Two-banded NH models}
\label{sec:two_band_NH}
The conceptually simplest framework for understanding most of the topological properties of NH band structures, including the occurrence of exceptional degeneracies in momentum space as well as the role of important symmetries, is provided by two-banded systems. We hence proceed by considering NH model Hamiltonians, which in reciprocal space at lattice momentum $k$ are of the generic form
\begin{equation}
H(k) = {\bf d}(k) \cdot \boldsymbol\sigma + d_0(k) \sigma_0,
\label{eqn:nhtwoband}
\end{equation}
where ${\bf d} = {\bf d}_R + i {\bf d}_I$ with ${\bf d}_R, {\bf d}_I \in \mathbb{R}^3$, $d_0 \in \mathbb{C}$, $\boldsymbol\sigma$ the vector of standard Pauli matrices, and $\sigma_0$ the $2 \times 2$ identity matrix. The complex-energy spectrum then explicitly reads as
\begin{equation}
E_\pm = d_0 \pm \sqrt{d_R^2 - d_I^2 + 2 i {\bf d}_R \cdot {\bf d}_I}, 
\label{eq:generic_eigenvalues_2x2}
\end{equation}
where we drop the $k$ dependence of all quantities for brevity.

\begin{figure}
  \includegraphics[width=0.9\columnwidth]{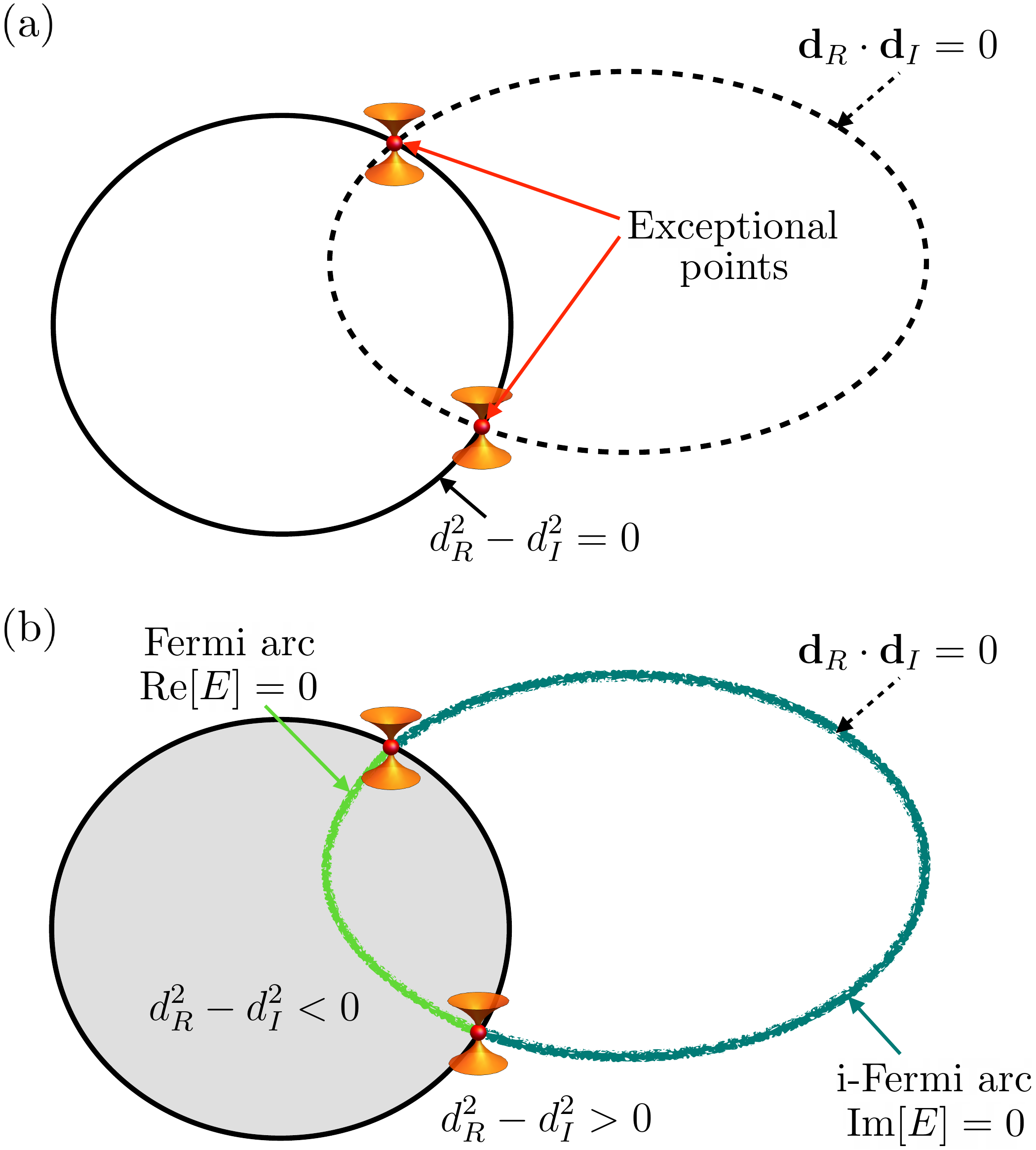}
  \caption{(a) Solutions to $d^2_R - d_I^2 = 0$ (solid line) and ${\bf d}_R \cdot {\bf d}_I=0$ (dashed line) [cf. Eq.~(\ref{eqn:ep_conditions})] form closed loops in a two-dimensional parameter space. Exceptional points appear when both equations are satisfied simultaneously, i.e., when the two loops intersect. (b) Exceptional points are connected by (imaginary) Fermi arcs: When ${\bf d}_R \cdot {\bf d}_I=0$ and $d^2_R - d_I^2 < 0$ (gray region), $\textrm{Re}[E] = 0$ [green (light gray) line], while $\textrm{Im}[E] = 0$ [dark green (dark gray)] when ${\bf d}_R \cdot {\bf d}_I=0$ and $d^2_R - d_I^2 > 0$ (outside the gray region).}
  \label{fig:generic_eps}
\end{figure}

{\emph{Abundance of exceptional degeneracies.}}---For Hermitian systems (implying ${\bf d}_I=0$), degeneracies in the spectrum (\ref{eq:generic_eigenvalues_2x2}) occur only if all three components of ${\bf d}_R$ are simultaneously tuned to zero. This is the basic reason why topologically stable nodal phases such as Weyl semimetals occur in three spatial dimensions in conventional band structures. However, allowing for ${\bf d}_I \ne 0$ in NH systems, from Eq.~(\ref{eq:generic_eigenvalues_2x2}) we see that degeneracies occur when
\begin{equation}
d_R^2 - d_I^2 = 0, \qquad {\bf d}_R \cdot {\bf d}_I = 0
\label{eqn:ep_conditions}
\end{equation}
are satisfied simultaneously, i.e., upon satisfying only two real conditions \cite{Berry2004}. This implies that nodal points in an NH band structure are generic and stable in two spatial dimensions, as shown schematically in Fig.~\ref{fig:generic_eps}.

\begin{figure*}
 \includegraphics[width=0.9\textwidth]{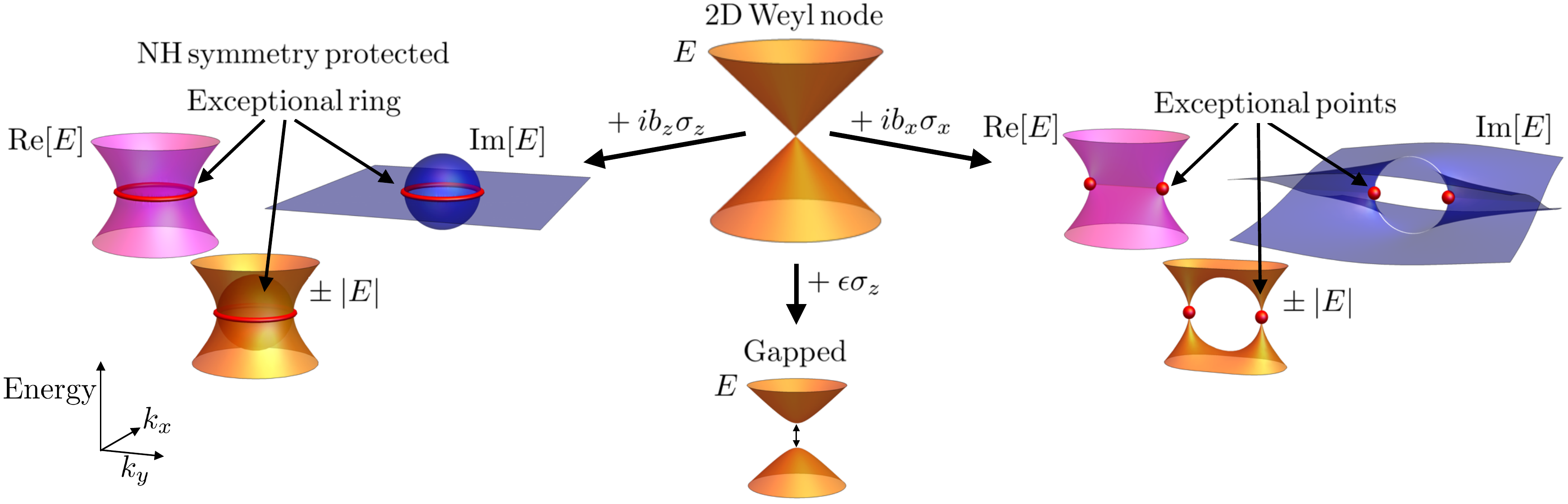}
  \caption{Schematic plot of the energy spectrum of a Hermitian, two-dimensional Weyl node $H = k_x \sigma_x + k_y \sigma_y$. Upon adding a mass term $\epsilon \sigma_z$, a gap opens in the spectrum (here shown for $\epsilon=0.1$). When instead an imaginary term $i b_z \sigma_z$ is added to the Hamiltonian, a ring of exceptional points, i.e., an exceptional ring appears (here shown in red for $b_z = 0.3$). The addition of an imaginary term $i b_x \sigma_x$ leads to the appearance of exceptional points (here shown in red for $b_x = 0.3$). The orange plots represent the absolute value of the energy $\pm |E|$, which for the Hermitian case simply corresponds to the energy $E$, whereas the pink and blue plots show the real and imaginary parts of the energy $\textrm{Re}[E]$ and $\textrm{Im}[E]$, respectively.}
  \label{fig:2D_WN}
\end{figure*}

Another key difference to Hermitian systems is that any nontrivial solutions to Eq.~(\ref{eqn:ep_conditions}) lead to degeneracies in the form of exceptional points, where the NH Hamiltonian becomes defective since the two eigenstates coalesce (become linearly dependent) upon approaching the degenerate eigenvalue. This is not the case for the trivial solution $ {\bf d}_R = {\bf d}_I = 0$, known as the diabolic point. The diabolic point concurs with the aforementioned Hermitian degeneracy condition, but has a much lower abundance as it requires fine-tuning of six parameters in the NH context. These simple algebraic observations on NH matrices have profound implications on the topological classification and physical properties of NH systems, which is elaborated on in Sec.~\ref{sec:nodalphases}.

\subsection{Nodal phases}
\label{sec:nodalphases} 
A natural question that has recently been the subject of intense theoretical and experimental study addresses to what extent the paramount algebraic phenomenon of EPs affects the physical properties of NH systems. In this section, we review recent results along these lines regarding both the topological classification and the physical phenomenology of NH band structures exhibiting EPs.

\subsubsection{Topological non-Hermitian metals}
\label{sec:nh_metals}
We illustrate the stable occurrence of NH nodal points in two spatial dimensions by perturbing a Hermitian two-dimensional (2D) Weyl point described by the model Hamiltonian
\begin{align}
H(k) = k_x \sigma_x + k_y \sigma_y
\label{eqn:2DHermitianWeyl}
\end{align}
in a NH fashion in various ways. The Hermitian perturbation $\epsilon \sigma_z$ is readily seen to immediately open a gap of the order of $\epsilon>0$ [see Eq.~(\ref{eq:generic_eigenvalues_2x2}) and Fig.~\ref{fig:2D_WN}], demonstrating the fine-tuned character of a 2D Weyl point in the Hermitian realm. By contrast, if we add the corresponding anti-Hermitian perturbation $i b_z \sigma_z,~b_z \in \mathbb R$ from plugging ${\bf d}_R=(k_x,k_y,0),~{\bf d}_I=(0,0,b_z)$ into Eq.~(\ref{eqn:ep_conditions}), we find a ring of exceptional degeneracies at $k^2=b_z^2$. That is, the system remains gapless; see Fig.~\ref{fig:2D_WN}. However, when considering the combination of these two perturbations, Eq.~(\ref{eqn:ep_conditions}) amounts to $k^2+\epsilon^2= b_z^2,~b_z \epsilon=0$, meaning that there is a gap as soon as both $\epsilon$ and $b_z$ are finite, thus rendering the aforementioned nodal ring unstable. More precisely, in Sec.~\ref{sec:nodal_spt}, we discuss the fact that such nodal structures of higher dimensions are stable only in the presence of certain NH symmetries. 

Next we choose an anti-Hermitian term $i b_x \sigma_x$, which at $\epsilon=0$ gives rise to degeneracies when $k^2 = b_x^2$ and $b_x k_x=0$, i.e., at the isolated points $(k_x,k_y)=(0,\pm b_x)$ (Fig.~\ref{fig:2D_WN}). In contrast to the ring degeneracy, these isolated EPs are stable against $\epsilon>0$, and for that matter against any small NH perturbation. More specifically, the isolated EPs will continuously move in momentum space as a function of generic perturbations and can be removed only if they meet in momentum space. This renders NH 2D systems with isolated nodal points in the form of EPs a topologically stable phenomenon defining a {\textit{NH Weyl phase}}. On a more formal note, as mentioned in Sec.~\ref{sec:intro}, the complex-energy spectrum at an isolated second-order EP behaves like a complex square-root function around the origin. Hence, such EPs in two dimensions form branch points in energy that can be removed only by contracting the branch cut connecting them.

An important physical consequence of the concomitant phase winding of the complex energy around the EPs is the existence of contours with purely imaginary (purely real) energy emanating from them, also called NH Fermi arcs (imaginary NH Fermi arcs, or $i$-Fermi arcs), which are equivalent to the aforementioned branch cuts; wee Fig.~\ref{fig:generic_eps}(b)\cite{koziifu,Zhou2018,carlstroembergholtz,Yang2019a}. While in our simple continuum model such contours can extend to infinite momenta, the compact nature of reciprocal space (the first Brillouin zone) in Bloch bands describing crystalline structures strictly enforces them to form open arcs connecting the EPs, which is somewhat reminiscent of Fermi arcs in conventional 3D semimetals. However, a crucial difference to their Hermitian counterpart is that NH Fermi arcs are a bulk phenomenon (similar in this regard to standard Fermi surfaces), while the surface Fermi arcs in 3D Weyl semimetals connect the projection of the Weyl points to a given surface \cite{Armitage2018}. Thus, 2D NH Weyl phases distinguished by the number of pairs of EPs are a NH counterpart of {\textit{metallic}} dispersions in solids, while generic Hermitian 3D Weyl systems represent {\textit{semimetallic}} band structures in this solid-state context.

\begin{figure}[t]
  \includegraphics[width=0.9\columnwidth]{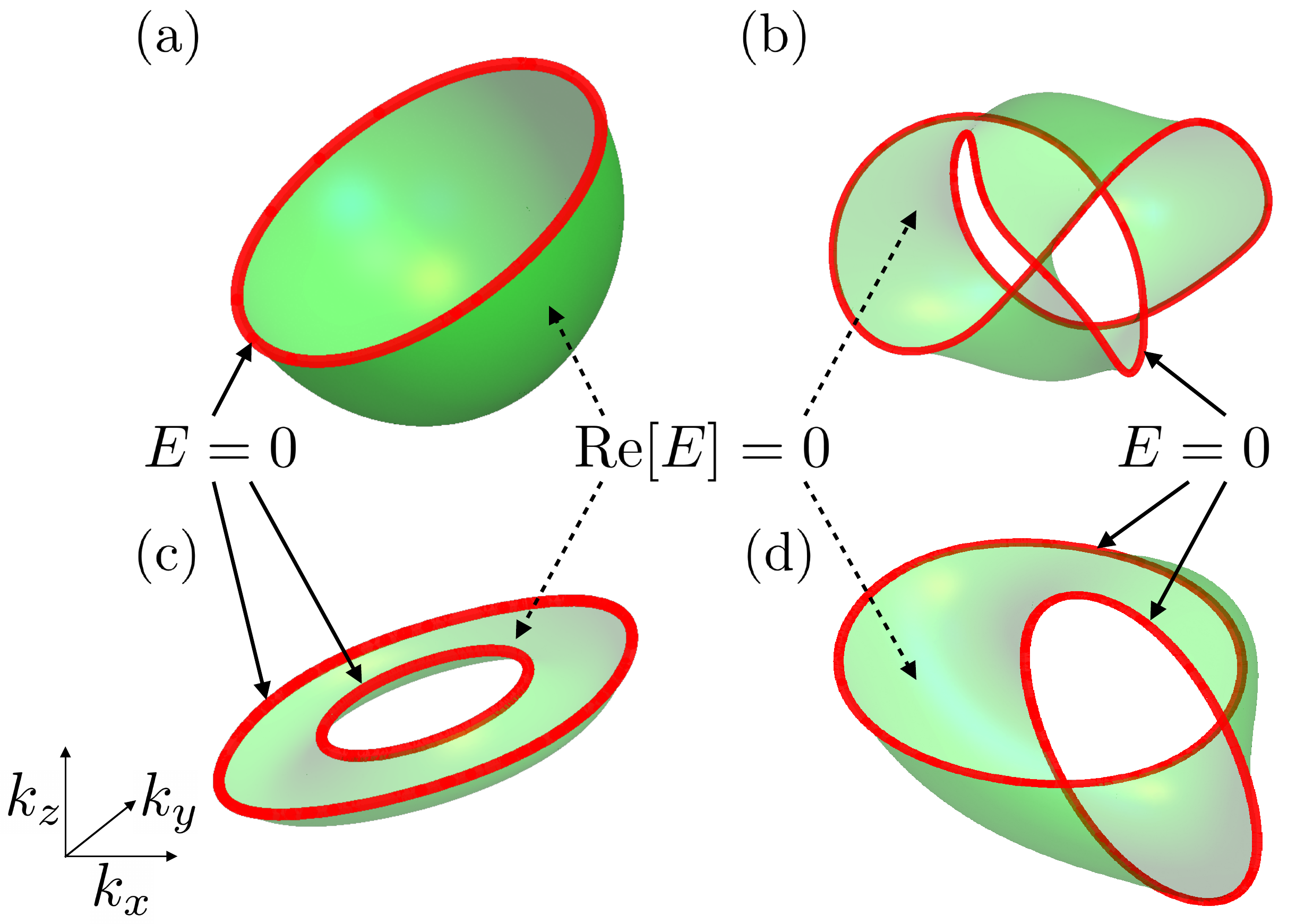}
  \caption{Exceptional rings or knots where the energy is degenerate [in red (dark gray)] and Seifert surfaces where $\textrm{Re}[E] = 0$ [in green (light gray)] appearing in the spectra of short-range hopping models resulting in (a) an exceptional ring, (b) a trefoil knot, (c) two exceptional rings, and (d) a Hopf link. See \onlinecite{knots} for Hamiltonian details.}
  \label{fig:seifert_surface_example}
\end{figure}

\emph{Knotted non-Hermitian metals.}---Moving to three spatial dimensions, the simple parameter counting in Sec.~\ref{sec:two_band_NH} tells us that EPs in three dimensions generically, i.e., without relying on fine-tuning or symmetries, form closed nodal lines in reciprocal space rather than occurring at isolated points \cite{Xu2017,Cerjan2019}. This allows for a new category of topologically stable NH metallic phases where the nodal lines themselves represent topologically nontrivial objects such as links \cite{carlstroembergholtz,Yang2019} or knots \cite{knots, Stalhammar2019}. By slicing such 3D systems into layers of 2D systems in reciprocal space, the aforementioned argument on NH Fermi arcs may be readily generalized to the 3D case in the following sense: Exceptional nodal lines necessarily bound open NH Fermi surfaces, which for knotted nodal structures appear in the form of Seifert surfaces; see Fig.~\ref{fig:seifert_surface_example}. Not only are these phenomena mathematical possibilities of academic interest, but in fact simple microscopic tight-binding models within experimental reach have recently been shown to exhibit a variety of linked and knotted NH nodal structures \cite{knots, Stalhammar2019, Li2019,Yang2019c}. The phenomenon of knotted nodal NH band structures has no direct counterpart in Hermitian systems. There, owing to the higher codimension of nodal points, additional symmetries are necessary to stabilize knotted or linked nodal lines \cite{Bi2017}, and such fine-tuned nodal structures would not entail Fermi-Seifert surfaces. 

\subsubsection{Symmetry-protected nodal phases}
\label{sec:nodal_spt}
Requiring symmetries is well known to generally refine a topological classification by constraining the set of eligible physical systems. Concretely, two model Hamiltonians that would be considered equivalent in the absence of a given symmetry may become distinct in its presence if any path adiabatically connecting them necessarily breaks that symmetry. This phenomenon defines the notion of symmetry-protected topological (SPT) phases \cite{Chen2013,ChiuRev}. 

{\textit{Symmetries in Hermitian systems.}}---In conventional Hermitian systems, a primary example of nodal SPT phases is provided by Dirac semimetals. There the spin-degenerate Dirac points may be continuously removed individually unless protecting symmetries such as the combination of parity and time-reversal symmetry (TRS) are postulated. This is in contrast to Weyl semimetals, the individual Weyl points of which are topologically stable without symmetries other than the lattice momentum conservation defining the Bloch band structure. 

A comprehensive symmetry classification was achieved in a seminal paper by Altland and Zirnbauer (AZ) \cite{Altland1997}. The AZ classification is based on generic symmetry constraints characterizing ensembles of mesoscopic systems beyond standard unitary symmetries that commute with the system Hamiltonian. Specifically, the considered constraints are the antiunitary TRS defined by\footnote{We state the symmetry constraints in Eqs.~(\ref{eqn:AZ_TRS})-(\ref{eqn:AZ_CS}) for a free Hermitian Hamiltonian in first-quantized form, on which the action of transposition and complex conjugation are equivalent.}
\begin{align}
T_\pm H^*T_\pm^{-1}=H, \qquad T_\pm T_\pm^* = \pm 1,
\label{eqn:AZ_TRS}
\end{align}
where the asterisk denotes complex conjugation, the particle-hole constraint (PHC)
\begin{align}
C_\pm H^* C_\pm^{-1} = - H,\qquad C_\pm C_\pm^* = \pm 1,
\label{eqn:AZ_PHC}
\end{align}
and, resulting from the combination of TRS and PHC, the chiral symmetry (CS)
\begin{align}
U_C H U_C^\dag = -H,\qquad U_CU_C^\dag = U_C^\dag U_C= U_C^2 = 1. \label{eqn:AZ_CS}
\end{align}
Considering all independent combinations of these constraints gives rise to the ten AZ symmetry classes, on the basis of which the periodic table of topological insulators was constructed \cite{SchnyderClassification2008, Kitaev2009, Ryu2010}. Later on, also considering conventional commuting unitary symmetries such as crystalline symmetries resulted in the identification of a plethora of additional (both gapped and nodal) topological band structures \cite{Fu2011, Ando2015,ChiuRev}.

{\textit{Generic symmetries in non-Hermitian systems.}}---The natural question of how the AZ symmetry classification may be generalized to NH systems was addressed by Bernard and LeClair (BLC) \cite{Bernard2002}, who derived a 43-fold symmetry classification for ensembles of NH matrices. This system of symmetries was proposed for the topological classification of bosonic Bogoliubov--de\,Gennes Hamiltonians by \onlinecite{LieuSym}. Here we review key elements of the general BLC classification and its recently proposed amendments \cite{Kawabata2018}, focusing on qualitative differences to the AZ classification in Hermitian systems. In essence, the main complication in NH systems relative to the Hermitian realm is that transposition ($H\rightarrow H^T$) and complex conjugation ($H\rightarrow H^*$) are inequivalent operations, and even Hermitian conjugation ($H \rightarrow H^\dag$) may act nontrivially on a given NH effective Hamiltonian $H$; see \cite{Kawabata2018} for a detailed discussion along these lines. As a consequence, both TRS and PHC split into two inequivalent NH generalizations, distinguished by whether or not complex conjugation is replaced by transposition in Eqs.~(\ref{eqn:AZ_TRS}) and (\ref{eqn:AZ_PHC}), respectively. Furthermore, the nontrivial action of Hermitian conjugation gives rise to so-called pseudo-Hermiticity constraints \cite{Mostafazadeh2002}
\begin{align}
Q_{\pm} H^\dag Q_{\pm}^{-1}= \pm H,\quad Q_{\pm}Q_{\pm}^\dag = Q_{\pm}^\dag Q_{\pm} = 1,
\label{eqn:PseudoHSym}
\end{align} 
where $Q_+$ ($Q_-$) are ordinary commuting (chiral anticommuting) symmetry constraints for Hermitian $H$, but gives rise to new symmetry classes in the generic NH case. Note that symmetries involving Hermitian conjugation leave the (quasi)momentum $k$ invariant, and thus lead to local constraints in reciprocal space, which change the codimension of the EPs in the complex spectra of Bloch Hamiltonians (as discussed later).

Since an additional minus sign upon complex conjugation may be generated simply by multiplication by the imaginary unit ($H \rightarrow i H$), TRS and PHS as defined in Eqs.~(\ref{eqn:AZ_TRS}) and (\ref{eqn:AZ_PHC}) may be mapped onto one another by considering $i H$ instead of $H$ \cite{Kawabata2019a}. The identification of these operations for classification then is, at least at a formal level, justified by the fact that the spaces of eligible Hamiltonians differing by a prefactor of $i$ are isomorphic. However, since physically a multiplication by $i$ has quite dramatic effects, it is fair to say that in real models these two cases may still correspond to quite different scenarios; see also Sec.~\ref{phys_constraints}. Taking into account all aforementioned symmetry constraints and relations, a counting of all independent symmetry classes leads to a grand total of 38~\cite{Kawabata2018}, rather than the 43 symmetry classes originally proposed by BLC.

{\textit{NH symmetries and abundance of EPs.}}---Given this NH symmetry classification, we now review and illustrate the effect of NH symmetries on the occurrence and stability of exceptional nodal structures in NH band structures \cite{Budich2019}. As discussed in Sec.~\ref{sec:two_band_NH}, in the absence of symmetries EPs have codimension 2 [see Eq.~(\ref{eqn:ep_conditions})] and thus generically appear at isolated points in 2D NH band structures and as closed lines in 3D NH band structures. 

Some basic intuition about how NH symmetries change this behavior can be gained again by considering two-banded models as introduced in Sec.~\ref{sec:two_band_NH} preserving the symmetry $Q_+$. For concreteness, we make the explicit choice $Q_+ = \sigma_x$. Then, the symmetry (\ref{eqn:PseudoHSym}) in Eq.~(\ref{eqn:nhtwoband}) implies the constraint ${\bf d}_R = (d_R^x,0,0),~{\bf d}_I = (0,d_I^y,d_I^z)$ which trivializes one of the conditions, namely ${\bf d}_R \cdot {\bf d}_I=0$ [see Eq.~(\ref{eqn:ep_conditions})], for obtaining EPs. Thus, the codimension of exceptional degeneracies is reduced from $2$ to $1$. As an immediate consequence, EPs at isolated points appear in 1D, and closed lines of EPs occur in 2D (see, e.g., Fig.~\ref{fig:2D_WN} for the appearance of an exceptional ring). This dimensional shift promotes the aforementioned bulk Fermi arcs to open Fermi volumes, as the surfaces bounded by the EPs now have the same spatial dimension as the system itself. 

This phenomenology is not limited to the minimal two-band setting at hand but has been shown to generalize to generic NH band structures in numerous BLC classes that contain reality constraints on the complex spectrum \cite{Budich2019}. A $K$-theory-based classification of gapless nodal NH phases was recently reported on by \onlinecite{Kawabata2019}, and we thus arrive at periodic tables encompassing all $38$ symmetry classes, as proposed by \onlinecite{Kawabata2018}. Instructive examples starting from four-band Dirac models were worked out explicitly by \onlinecite{Rui2019a} and symmetry-protected rings of EPs are know to naturally emerge in honeycomb-based systems \cite{Yoshida2019a,Szameit2011}.

\subsubsection{Higher-order exceptional points} \label{sect:higher_order_eps}
We now discuss the existence of higher-order EPs, which we encountered in our discussion of the Hatano-Nelson model; see Sect.~\ref{sec:one_band_NH}. In multiband systems, an EP of the order of $n$ appears when the Hamiltonian matrix features an $n$-dimensional Jordan block $J$ as in the following:
\begin{equation}
J = \begin{pmatrix}
E & 1 & 0 & 0 & 0 \\
0 & E & 1 & 0& 0 \\
0 &0 & E & \cdots & 0\\
0 & 0 & \vdots & \ddots & 1 \\
0 &0& 0 & 0 & E
\end{pmatrix},\label{eq:jordanblock}
\end{equation}
with $E$ the eigenvalue of the EP. As the Hamiltonian matrix may feature multiple such Jordan blocks of varying dimensions, EPs of different orders can coexist in the band spectrum. To find an $n$th-order EP one needs to tune $2n-2$ parameters \cite{Holler2018}, such that the appearance of EPs of higher order requires an increasing amount of parameter fine-tuning. Thus, largely unexplored topological nodal phases featuring EPs of the order of $n$ are readily predicted to generically occur in $d=2n-2$ dimensions. 

\begin{figure*}
  \includegraphics[width=0.8\textwidth]{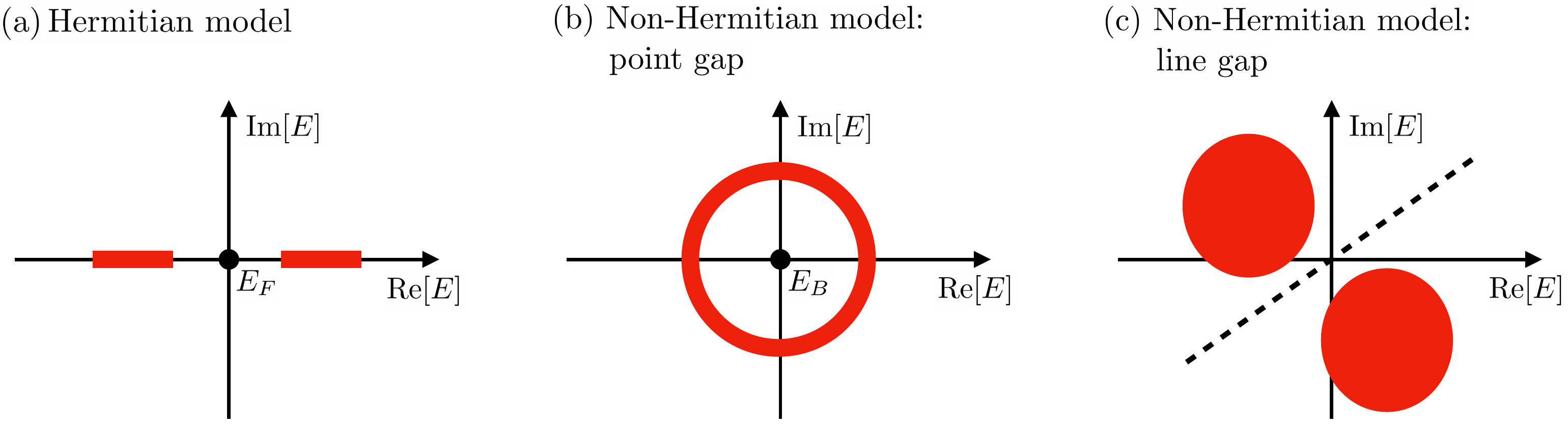}
  \caption{Schematic depiction of (a) gaps in Hermitian models, (b) point gaps and (c) line gaps in NH models, with the bulk bands shown in red (gray).}
  \label{fig:point_line_gap}
\end{figure*}

Perturbing around an EP of the order of $n$ with $\omega$ generally leads to the Puiseux series, $E \approx E_0 + \omega^{1/n} E_1 + \omega^{2/n} E_2 + \mathcal{O}(\omega^{3/n})$, which means that an $n$th-order EP scales with the $n$th root. \onlinecite{Demange2011} showed, however, that not all higher-order EPs scale in this fashion. For example, a third-order EP may feature square-root behavior \cite{Demange2011}. In Sec.~\ref{sec:NH_BBC} we discuss that even though EPs with high order are in principle rare in the space of all models, they readily appear in the open-boundary-condition spectrum of models that break conventional bulk-boundary correspondence due to their close relation with the NH skin effect. A simple example of this can be observed for the Hatano-Nelson model with unidirectional hopping; cf. Eq.~(\ref{eq:hatano_nelson_model}), which at $J_L=0$ or $J_R=0$ takes the form of Eq.~(\ref{eq:jordanblock}) for open boundary conditions.

\subsection{Gapped phases}
\label{sec:gapped}
We now turn to the topological classification of gapped NH systems, again focusing on crucial differences to the conventional Hermitian realm, where the periodic table of topological insulators and superconductors based on the AZ symmetry classification by now has become a widely known amendment to the theory of Bloch bands.

\subsubsection{Point gaps versus line gaps}
\label{sec:point_vs_line}
The first crucial observation when moving to NH band structures with complex-energy spectra is that there is no canonical way of defining a spectral gap. To overcome this issue, \onlinecite{Kawabata2018} recently proposed to classify complex-energy gaps into two categories: point and line gaps. A NH model is said to have a point gap when the complex-energy bands do not cross a base point $E_B$, and where crossing this base point defines a gap closing transition; wee Fig.~\ref{fig:point_line_gap}. A line gap, on the other hand, is defined by a line in the complex-energy plane, which has no intersections with the energy bands; see Fig.~\ref{fig:point_line_gap}. Note that models with a line gap also always have a point gap. Line gaps in complex spectra carry close similarities to energy gaps in Hermitian models \cite{Kawabata2018}, as a spectrum of a Hermitian model is said to be gapped when there are no energy bands that cross the Fermi energy $E_F$. Indeed, in both the Hermitian and NH case, the individual bands in a spectrum with a line gap can be contracted to single points. Point gaps that do not generalize to line gaps do not have a direct Hermitian counterpart and are thus genuinely non-Hermitian. 

Recently it was shown that certain $d$-dimensional NH models with a point gap can be naturally interpreted as the ``surface theory" of $(d+1)$-dimensional, Hermitian models \cite{Lee2019,gong,FoaTorres2019,Terrier2020}, where these models are formally related via a doubling procedure and dimensional ascension or reduction \cite{Lee2019}. Following \onlinecite{Lee2019}, this relation may be intuitively understood at the level of the Hatano-Nelson model [see Eq.~(\ref{eq:hatano_nelson_model})]: In the long-time limit, there is only one chiral mode in the system. Indeed, at $\textrm{Re}(E_k) = 0$, i.e., $k = \pm \pi/2$, it is possible to find two modes with opposite chirality: one mode with group velocity $v_{\pi/2} = \left.\textrm{Re} \left(\partial_k E_k \right)\right|_{k = \pi/2} = - \left(J_L + J_R\right)$, and another mode with $v_{-\pi/2} = J_L + J_R$. The lifetime of these modes is set by $\textrm{Im}(E_k)$, which is found to be positive (negative) for the mode with group velocity $v_{\pi/2}$ ($v_{-\pi/2}$). In the long-time limit, only the mode with $\textrm{Im}(E_k)>0$ survives, such that we are left with a single chiral mode. In this sense, this one-dimensional non-Hermitian model realizes the anomalous edge behavior of the two-dimensional quantum Hall effect, and can thus be interpreted as the ``edge theory'' of the latter \cite{Lee2019}.

\subsubsection{Symmetry-protected point-gapped phases}
\label{sec:point_gapped_spt}
The base energy $E_B$ with respect to which a point gap is defined may without loss of generality be chosen as $E_B=0$, which at most amounts to adding a constant complex-energy shift to a given Hamiltonian. Then the set of all admissible NH Bloch Hamiltonians is simply given by the general linear group formed of all regular complex matrices GL$(n, \mathbb C)$, where $n$ is the number of bands. Without additional symmetries, the set of inequivalent strong topological NH phases in $d$ spatial dimensions for $n>d/2$ is then given by 
\begin{align}
\pi_d\left[\textrm{GL}(n, \mathbb C)\right]=\begin{cases} \mathbb Z,~d \text{ odd,}\\0,~d \text{ even,}\end{cases}
\end{align}
i.e., by the $d$th homotopy group of GL$(n, \mathbb C)$ \cite{SchnyderClassification2008, Budich2013}. Non-symmetry-protected topological NH band structures thus occur in odd spatial dimensions \cite{gong}, in stark contrast to Hermitian systems, where the $m$th Chern number in $d=2m$ characterizes topological band structures that do not rely on additional symmetries \cite{Ryu2010}. For the simplest conceivable case $d=n=1$, the explicit invariant characterizing a given model Hamiltonian is given by the spectral winding number defined in Eq.~(\ref{eqn:spectralwinding}). This can be generalized to an arbitrary $n>1$ by simply replacing $E_k \rightarrow \text{det} H(k)$, and to odd $d>1$ as a standard higher-dimensional analog of the winding number known from chiral symmetric systems in the Hermitian realm; see Eq.~(\ref{eq:winding_number_generic}) \cite{Ryu2010}. This correspondence is not a coincidence, and it was shown by \onlinecite{gong} that any NH Hamiltonian $H$ may be augmented by a CS-preserving Hermitian Hamiltonian
\begin{align}
\tilde H=\begin{pmatrix}0&H\\ H^\dag&0\end{pmatrix} \label{eq:doubled_hamiltonian}
\end{align}
acting on a doubled Hilbert space, such that the standard Hermitian chiral invariant associated with $\tilde H$ concurs with the NH spectral winding invariant 
\begin{align}
w_{2n+1} &= \frac{(-1)^n n!}{(2n+1)!}\left(\frac i {2\pi}\right)^{n+1}\epsilon^{\alpha_1\alpha_2\cdots} \nonumber \\
&\times \int_{{\rm BZ}} {\rm tr}[H^{-1}(\partial_{k_{\alpha_1}} H) \cdot H^{-1}(\partial_{k_{\alpha_2}}H)\cdots] d^{2n+1} k \label{eq:winding_number_generic}
\end{align}
in $d=2n+1$ dimensions \cite{SchnyderClassification2008, Budich2013}.
Based on these observations and the AZ symmetry classification \cite{Altland1997} (see also Sec.~\ref{sec:nodal_spt}), \onlinecite{gong} arrived at a first NH counterpart of the periodic table of topological insulators. However, as discussed in more detail in Sec.~\ref{sec:nodal_spt}, the nontrivial action of Hermitian conjugation in NH systems naturally refines the tenfold AZ classification to the 43-fold BLC classes, later proposed to be reducible to a 38-fold way \cite{Kawabata2018}. Adapting the $K$-theory methods used by Kitaev \cite{Kitaev2009} for the Hermitian periodic table to this NH scenario, topological classification tables for gapped phases based on the BLC symmetry classification have recently been derived \cite{Kawabata2018, ZhouPeriodicTable}.

\subsubsection{Symmetry-protected line-gapped phases}
\label{sec:line_gapped_spt}
Regarding gaps in the shape of a straight line in the energy spectrum, in principle any offset and orientation in the complex plane may be considered to start with. However, as in the point-gapped case, by means of a constant energy shift, the gap line may be transformed to cross the origin. Furthermore, by rescaling the Hamiltonian with a complex constant, such a gap line may then be rotated to, say, the real energy axis. Since such a rotation of the energy spectrum may violate, or at least transform, generic NH symmetries, \onlinecite{Kawabata2018} still distinguished line gaps along the real and imaginary axis due to their distinct behavior under spectral reality constraints. 

For the case of a real line gap, any NH model Hamiltonian may be continuously deformed into a Hermitian Hamiltonian without breaking of symmetries, which reduces the classification problem to that of Hermitian matrices. In the case of an imaginary gap a similar deformation to an anti-Hermitian Hamiltonian $H_a$ is always possible. However, since the NH symmetries may be transformed in a nontrivial way when rotating to the Hermitian Hamiltonian $iH_a$, the classification problem of NH Hamiltonians with imaginary line gaps amounts to that of Hermitian systems up to a shift in symmetry class. Based on these observations, periodic tables for line-gapped Hamiltonians in all $38$ symmetry classes were obtained by \onlinecite{Kawabata2018}. Furthermore, \onlinecite{Liu2019a} considered the classification of defects in the BLC classes and generalizations thereof.

\subsection{Complementary classification approaches}
\label{phys_constraints}
Thus far our discussion of NH topological band structures has been based on the BLC symmetry classification, which is a direct NH generalization of the celebrated AZ classification of electronic systems in the Hermitian realm. Given the broad spectrum of applications of effective NH Hamiltonians (see Sec.~\ref{sec:NH_applications} for an overview), depending on the given physical situation, differing from the BLC classification by considering other symmetries and physical constraints can be natural. In the following, we briefly highlight some prominent examples of deviations from the classification discussed in Secs.~\ref{sec:nodalphases} and \ref{sec:gapped}.

\subsubsection{Other symmetries}

The combination of time-reversal symmetry and parity, widely known as $PT$ symmetry, was originally considered a fundamental NH amendment to quantum physics \cite{Bender1998}, as it gives rise to reality constraints on the spectrum known as pseudo-Hermiticity \cite{Mostafazadeh2002}, similar to the aforementioned constraint $Q_+$ [see Eq.~(\ref{eqn:PseudoHSym})] from the BLC system of symmetries. By now, $PT$-symmetry is widely established in effective NH descriptions of a variety of physical settings including photonic systems \cite{Regensburger2012, Ozdemir2019, Feng2017, Zyablovsky2014, yuce}. In particular, in the context of NH topology, states protected by $PT$-symmetry had been observed in optical systems \cite{NHexp} even before the systematic classification of NH symmetry-protected topological phases (see Secs.~\ref{sec:nodalphases} and \ref{sec:gapped}) was reported. As a second example outside of the BLC classification, a loose analog of supersymmetry, considered in high energy as a fundamental amendment to the standard model, has been identified in certain optical settings \cite{Miri2013,Heinrich2015}. Moreover, \onlinecite{Liu2019b} have classified NH phases with reflection symmetry, while a topological classification beyond the Hermitian realm was presented for dynamically stable systems by \onlinecite{DeNittis2019}.

\subsubsection{Fundamental constraints in quantum many-body systems} The BLC symmetry classification applies to generic NH matrices. However, when NH Hamiltonians are employed to effectively describe some form of dissipation in quantum many-body systems, inherent physical constraints reduce the space of eligible matrices. For example, the spectrum of effective Hamiltonians derived from a retarded Green's function including a NH self-energy is constrained to lie in the lower complex half-plane ${\rm Im}[E]\leq 0$; see \onlinecite{Bergholtz2019} for a recent discussion in the context of NH topological phases. This immediately rules out spectral winding around the origin [cf. Eq.~(\ref{eqn:spectralwinding})] and vortices in the complex spectrum as discussed in Sec.~\ref{sec:one_band_NH}, thus directly affecting the topological classification. A similar constraint appears when considering Liouvillian operators governing the dynamics of open quantum systems as NH matrices \cite{LindbladSkin,Lieu2019}. The basic physical meaning of such constraints is that quantum dissipation can damp out energy eigenstates (negative imaginary part) or leave them decoherence free (zero imaginary part) but not amplify their weight, which would correspond to a positive imaginary part.

\subsubsection{Homotopy perspective} 
Finally, we note that from Hermitian systems it is well known that there are so-called fragile topological phases [for a recent discussion see, e.g., \onlinecite{Kennedy2016a}] that do not survive the addition of extra bands. Such phases are not captured by the $K$-theory approach of the previously described classification schemes, but they can be described within a homotopy-theory-based classification \cite{Kennedy2016, Kennedy2016a}. In the NH context, new fragile topological phases have recently been uncovered by analyzing NH band structures from the vantage point of homotopy
\cite{Li2019a,Wojcik2019}. It is worth noting that such fragile phases relying on a low number of bands even exist in the absence of additional symmetries.

\section{Anomalous bulk-boundary correspondence} \label{sec:NH_BBC} 

In this section, we review recent findings on a phenomenology unique to NH systems, namely, qualitative changes in the so-called bulk-boundary correspondence (BBC), a fundamental principle for topological phases \cite{HasanKane}. In conventional Hermitian systems, the BBC establishes a one-to-one relation between topological invariants defined for infinite periodic systems and protected gapless boundary states occurring in systems with open boundaries. By contrast, in NH topological systems the BBC in its familiar form is found to generically break down (see Sec.~\ref{sect:breakdown_conventional_bbc}), and qualitative amendments to reestablish a modified NH BBC have been proposed; see Sec.~\ref{sec:approaches_nh_bbc}. For clarity, the conventional BBC known from Hermitian systems is in the following referred to as cBBC. We note that even in cases where cBBC holds, the transitions between different topological phases may be different from Hermitian systems, as they happen via exceptional degeneracies rather than Hermitian band-touching points \cite{Comaron2020,KuEdBuBe2018}. While the following discussion focuses mostly on the conceptually simple example of one-dimensional systems, we stress that surface states in NH topological systems are by no means limited to spectrally isolated bound states, but instead may also appear in higher-dimensional systems, e.g., in the form of chiral modes in NH Chern insulator models \cite{KuEdBuBe2018,YaSoWa2018}.   

\subsection{Breakdown of the conventional bulk-boundary correspondence} \label{sect:breakdown_conventional_bbc}
In this section, we review the mechanisms that lead to the breakdown of cBBC, i.e., the failure of topological invariants computed from the Bloch Hamiltonian to correctly predict the existence of boundary states. Furthermore, we discuss the \emph{NH skin effect} as well as the \emph{spectral instability of NH matrices}, which accompanies the breakdown of cBBC. There is, however, no strict one-to-one relation with these phenomena since the skin effect can also occur in systems where the cBBC does not have a clear meaning, as in systems with only point gaps.

\begin{figure}
  \includegraphics[width=0.9\columnwidth]{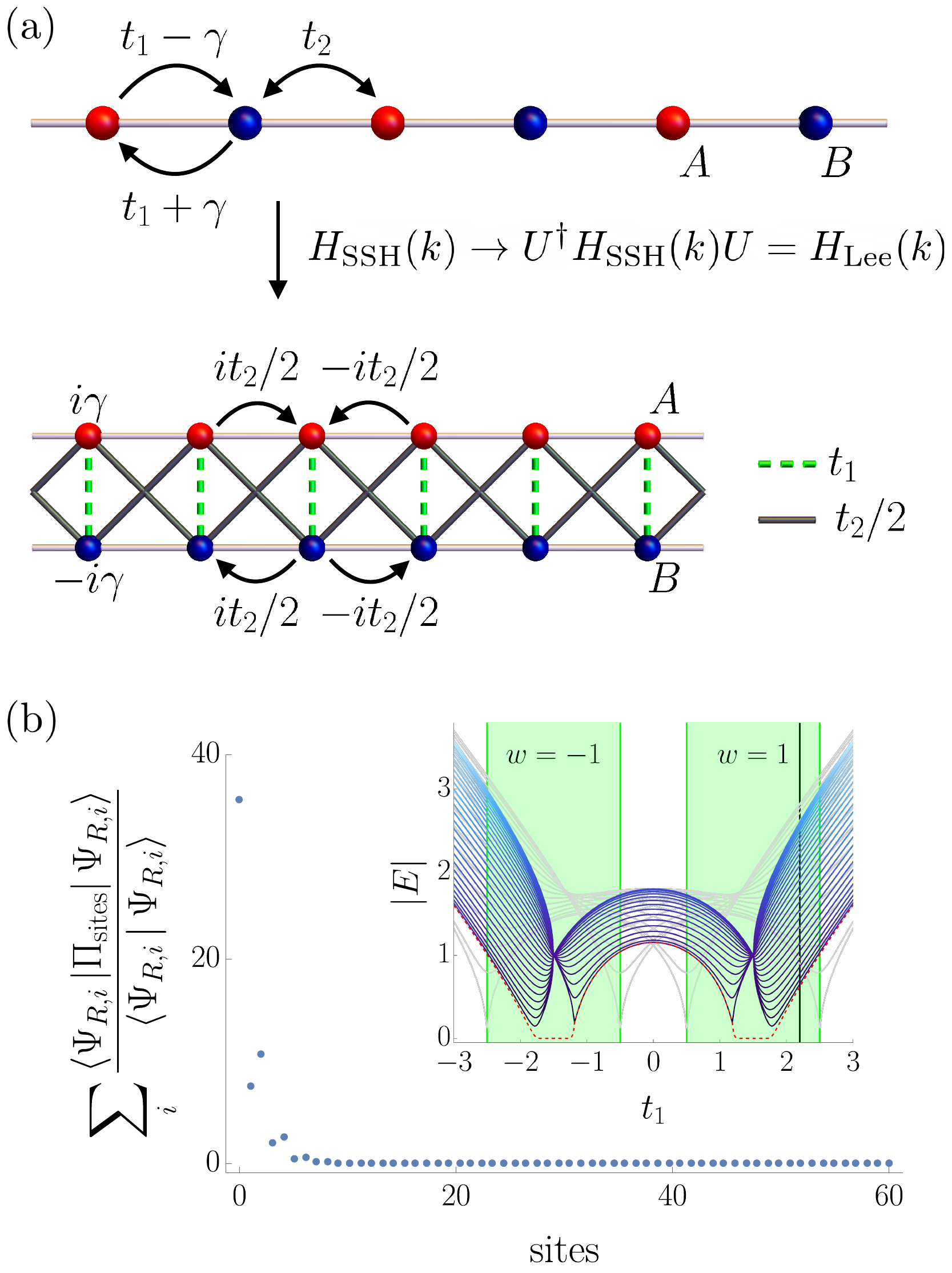}
  \caption{(a) Schematic depiction of the NH-SSH model (top panel) [see Eq.~(\ref{eq:nh_ssh_ham_real_space})] and the Lee model (bottom panel) \cite{LeeSkinEffect2016} and their unitary equivalence. (b) Sum of absolute squares of amplitudes per site of all right eigenstates for the Hamiltonian in Eq.~(\ref{eq:nh_ssh_ham_real_space}) with OBCs for $t_1 =2.2$, $t_2 = 1$, $\gamma = 1.5$, and $30$ unit cells. For this choice of parameters the magnitude of hopping to the left ($t_1 + \gamma = 3.7$) is larger than hopping to the right ($t_1 - \gamma = 0.7$), and we observe a piling up of states at the left end. Inset: absolute value of the eigenvalues as a function of $t_1$ for the same parameter choice with OBCs and PBCs in blue (dark gray) and light gray, respectively, and with the in-gap end states in the OBC case in red (dashed lines). The nonzero value of the winding number is explicitly indicated by green shaded areas and the black line corresponds to the value of $t_1$ for which the wave-function localization is plotted. We note that because the PBC and OBC Hamiltonians for the NH-SSH model and Lee's model are related via a unitary transform $U_N$, the PBC and OBC spectra, respectively, are identical for Lee's model. }
  \label{fig:abs_spec_nh_ssh_mod}
\end{figure}

\subsubsection{Canonical models and their interrelation} \label{subsubsec:canonical_examples}
The breakdown of cBBC in NH models was first observed by \onlinecite{LeeSkinEffect2016}, where a Creutz ladder with complex hopping terms and onsite dissipation [see the bottom panel of Fig.~\ref{fig:abs_spec_nh_ssh_mod}(a)] was studied. This phenomenon may be attributed to the anomalous behavior of the bulk states that, in the case of open boundary conditions (OBC), \emph{pile up} at the boundaries \cite{Xiong2018, KuEdBuBe2018}; see also Sec.~\ref{subsubsec:skin_effect} for a more detailed discussion. The easiest and most intuitive way of breaking cBBC is by including hopping terms in the tight-binding Hamiltonian, whose tunneling strengths are direction dependent (anisotropic); see the upper panel of Fig.~\ref{fig:abs_spec_nh_ssh_mod}(a). As a consequence, the bulk states can propagate around the system in the preferred direction for periodic boundary conditions (PBCs), while they are found to pile up at the boundaries in the case of OBCs; see Fig.~\ref{fig:abs_spec_nh_ssh_mod}(b). This extreme difference in the behavior of the bulk states under different boundary conditions intuitively invalidates the authority of bulk topological invariants computed for PBCs in determining the existence of boundary states. To explicate and exemplify this exotic behavior, we start by studying a one-dimensional, conceptually simple example, which displays features similar to those reported by \onlinecite{LeeSkinEffect2016}. We consider a NH version \cite{Lieu} of the Su-Schrieffer-Heeger (SSH) chain \cite{SSH} as described by the Hamiltonian 
\begin{align}
&H_\textrm{SSH} = \sum_{n=1}^N \left[\left(t_1+\gamma\right) c^\dagger_{A, n} c_{B,n} + \left(t_1 - \gamma \right) c^\dagger_{B,n} c_{A,n} \right. \label{eq:nh_ssh_ham_real_space} \\
&\left. + t_2 \left(c^\dagger_{A, n+1} c_{B, n} + c^\dagger_{B, n} c_{A, n+1} \right)\right], \qquad t_1, t_2, \gamma \in \mathbb{R}, \nonumber
\end{align}
where $c^\dagger_{\alpha,n}$ ($c_{\alpha,n}$) creates (annihilates) a state on sublattice site $\alpha \in \{A, B\}$ in unit cell $n$, $N$ is the total number of unit cells, $t_1$ and $\gamma$ are the nearest-neighbor (NN) hopping parameters inside the unit cell, and $t_2$ is the NN hopping parameter between unit cells; see the top panel of Fig.~\ref{fig:abs_spec_nh_ssh_mod}(a) \cite{KuEdBuBe2018, YinJiangLiLuChen2018,YaWa2018}. Hermiticity is broken when $\gamma \neq 0$, which results in a different modulus of the hopping amplitude between hopping to the left with respect to hopping to the right inside the unit cell. The Bloch Hamiltonian is of the general form given in Eq.~(\ref{eqn:nhtwoband}), here with ${\bf d}(k) = \left(t_1 + t_2 \, \textrm{cos} k, \, t_2 \, \textrm{sin}k + i \gamma, \, 0 \right), \,d_0 (k) = 0$, where the presence of an imaginary anti-Hermitian term $i \gamma \sigma_y$ formally signals the breaking of Hermiticity. In the inset in Fig.~\ref{fig:abs_spec_nh_ssh_mod}(b), we plot the absolute value of the band spectrum for OBCs (in blue) and PBCs (in gray), observing a clear discrepancy. In this sense, the direction-dependent hopping is accompanied by a spectral instability; see Sec.~\ref{subsec:spectral_instability} for a more general discussion.

As with the Hermitian SSH chain, this model has a chiral symmetry, i.e., $\{H, \, \sigma_z\} = 0$, and it is thus possible to define a winding number, where in the Hermitian case this winding number determines the number of states localized to the ends \cite{RyuTopological2002, SchnyderClassification2008}. The nonzero values of the NH counterpart of the winding number \cite{gong,Kawabata2018}, i.e., the spectral winding number [cf. Eq.~(\ref{eqn:spectralwinding}) with $E_k$ replaced by $\textrm{det} H(k)$], are indicated explicitly in the spectrum in the inset in Fig.~\ref{fig:abs_spec_nh_ssh_mod}(b) by the green shaded areas. Here, unlike the conventional case, the winding number fails to predict the existence of the end states in the OBC case, which are shown in red in the OBC spectrum. In fact, the winding number changes value when a gap closing appears in the PBC spectrum, which is at strikingly different parameter values than when the OBC system features phase transitions. To further elucidate what is going on we plot the sum of the amplitude per site for all wave functions in the case of OBCs in Fig.~\ref{fig:abs_spec_nh_ssh_mod}(b), confirming that the wave functions indeed pile up at the boundary. In summary, the simple model defined by Eq. (\ref{eq:nh_ssh_ham_real_space}) indeed breaks cBBC and exhibits NH-skin-effect behavior, which was recently confirmed in several experiments \cite{GhBrWeCo2019,HeHoImAbKiMoLeSzGrTh2019, HoHeScSaetal2019, XiDeWaZhWaYiXu2019,Weidemann2020}.

We note that models in which the modulus of the hopping amplitudes is explicitly direction dependent, such as in Eq.~(\ref{eq:nh_ssh_ham_real_space}), are sometimes referred to as ``nonreciprocal hopping models'' in the literature \cite{HoHeScSaetal2019}, and the accumulation of bulk states at a boundary is often attributed to this property \cite{Lee2019b}. While this is analogous to nonreciprocal optical models, where the symmetry of wave transmission is broken [see \onlinecite{Sounas2017} for a review], the straightforward translation of this definition to the language of tight-binding models with internal degrees of freedom may not be unambiguous. When interpreting the sublattice degree of freedom of the NH-SSH model in Eq.~(\ref{eq:nh_ssh_ham_real_space}) as a spin rather than a spatial degree of freedom, the model would no longer be nonreciprocal in the aforementioned sense: While the internal coupling strengths between the two spins have a different magnitude, the hopping magnitude between lattice sites is no longer direction dependent. Nevertheless, in this differing interpretation, the model still exhibits all the aforementioned properties. We now demonstrate that the ambiguity of this notion of reciprocity goes much further.

In particular, a simple unitary transformation relates the Bloch Hamiltonians of the NH-SSH model and the Lee model
\begin{equation}
\!\!\!\! H_\textrm{SSH}(k)\! \rightarrow U^\dagger\! H_\textrm{SSH}(k) U\!\! = \! H_\textrm{Lee}(k), \ \ U \!=\! \frac{1}{\sqrt{2}}\!\begin{pmatrix} 1 & i \\ i & 1 \end{pmatrix}\!.
\end{equation}
Here we can directly identify ${\bf d}_\textrm{Lee}(k) = \left(t_1 + t_2 \, \textrm{cos} k, \, 0, \, t_2 \, \textrm{sin}k + i \gamma \right)$, $d_{0, \textrm{Lee}} (k) = 0$ 
for $H_\textrm{Lee}(k)$ \cite{LeeSkinEffect2016}. In this model it is natural to interpret $\gamma$ as on-site gain ($+ i \gamma$) and loss ($-i \gamma$), while $t_1$ and $t_2$ remain standard Hermitian NN hopping parameters inside and between unit cells, respectively; see the lower panel in Fig.~\ref{fig:abs_spec_nh_ssh_mod}(a). Moreover, also with OBCs it is easy to show that one may write $U_N = \mathbb{1}_N \otimes U$, where $\mathbb{1}_N$ is the identity matrix of dimension $N$, such that $U^\dagger_N H_\textrm{SSH}^\textrm{OBC} U_N = H_\textrm{Lee}^\textrm{OBC}$ with $U_N$ again being unitary. Thus, the spectra of Lee's model \cite{LeeSkinEffect2016} with either PBCs or OBCs are identical to those of the NH-SSH model, as shown in the inset of Fig.~\ref{fig:abs_spec_nh_ssh_mod}(b). It follows that the Lee model also exhibits a similar accumulation of bulk states at the boundary, since the unitary transformation $U_N$ acts only locally and hence does not drastically alter the localization of the eigenstates. 

In summary, while Lee's model [see the bottom panel of Fig.~\ref{fig:abs_spec_nh_ssh_mod}(a)] contains only \textit{diagonal on-site} gain and loss terms, it is related to a model with \textit{explicitly anisotropic hoppings} through a local unitary transformation. This observation further blurs the difference between ``reciprocal" and ``nonreciprocal" tight-binding models, as inferred from the symmetries of their hoppings, and we thus refrain from such a distinction in this review. Instead, we emphasize that the breakdown of the cBBC is a generic NH phenomenon not tied to a specific microscopic provenance of the non-Hermiticity.

\subsubsection{Non-Hermitian skin effect} \label{subsubsec:skin_effect}

The concept of a BBC relies on the doctrine that introducing boundaries into a model does not have significant effects on the bulk states, meaning that the model does not undergo a topological phase transition when going from PBCs to OBCs. In stark contrast, the behavior of the bulk states associated with the family of cBBC-breaking NH models studied in this section is altered in an extreme way upon considering OBCs: These models feature the NH skin effect [see Fig.~\ref{fig:abs_spec_nh_ssh_mod}(b)], a term coined by \onlinecite{YaWa2018}.

Intuitively, the appearance of the localized bulk states, which are also called \emph{skin states}, can be understood from the presence of or proximity to one or more high-order EPs (cf. Sec.~\ref{sect:higher_order_eps}), through which the states need to pass when tuning from PBCs to OBCs \cite{Xiong2018}. The appearance of these EPs, which scale with system size (infinite order EPs occur in the thermodynamic limit), similar to what we saw for the Hatano-Nelson model in Sec.~\ref{sec:one_band_NH}, results in a topological distinction between the model with PBCs and OBCs thus leading to a natural breaking of cBBC \cite{Xiong2018}. The connection between higher-order EPs, say, $n$th-order EPs at which $n$ eigenstates coalesce (cf. Sec.~\ref{sec:nodalphases}), and the piling up of bulk states can then be understood as follows: Close to such an EP, a macroscopic number $n$ of eigenstates necessarily have large spatial overlap, which is achieved through their accumulation at the same boundary.

This NH skin effect always appears when cBBC is broken and can thus be seen as a telltale signature thereof. The anomalous localization behavior of the bulk states does not find a counterpart in Hermitian physics and is thus an inherently NH phenomenon.

It is natural to ask which minimal ingredients are needed for a NH hopping model to possess skin states, and thereby to break cBBC. While not a sufficient criterion, a necessary requirement is that the Hermitian ($H_H = H_H^\dagger$) and anti-Hermitian [$iH_A = -(iH_A)^\dagger$] parts of the NH Hamiltonian $H=H_H + i H_A$ do not commute, i.e., $[H_H, \, H_A] \neq 0$.\footnote{An equivalent way of stating the necessary condition for skin states is that $H$ cannot be normal.} If they do commute, then $H_H$ and $H_A$ share a common eigenbasis, which means that the eigenstates of $H$ are the eigenstates of a \emph{Hermitian} matrix, namely, of $H_H$ (and $H_A$), and as a consequence, the corresponding eigenstates form a standard orthonormal basis and can as such not be skin states.

\onlinecite{Longhi2019a} showed that the existence of the NH skin effect in one-dimensional NH models can be detected by making use of a bulk probe: If the maximum value of the Lyapunov exponent in the long-time limit is at a drift velocity other than zero, this is a sufficient condition for the NH model to display the NH skin effect as well as symmetry-breaking phase transitions in the OBC spectrum.

A recent suggestion is that the presence of a topologically nontrivial point gap in the complex-energy spectrum of the Bloch Hamiltonian is equivalent to the eigenstates in the OBC system being skin states \cite{ZhangYangFang2019, OkumaKawabataShioSato2019}, which was also observed by \onlinecite{Wanjura2019}.

One may ask whether the piling up of states is forbidden by certain symmetries. Indeed, \onlinecite{Kunst2019} showed that $PT$-symmetric models (cf. Sec.~\ref{phys_constraints}) in the $PT$-unbroken phase \cite{Bender1998} cannot possess skin states, which was corroborated by \onlinecite{Kawabata2018}, who also showed that models in the presence of a parity inversion symmetry TRS$^\dagger$, which is defined as the relation in Eq.~(\ref{eqn:AZ_TRS}) with $H^* \rightarrow H^T$, or pseudo-Hermiticity in the unbroken phase (cf. Sec.~\ref{sec:nodal_spt}) are also excluded from exhibiting a breakdown of cBBC. It can be intuitively understood why these symmetries prevent the existence of skin states: For example, $PT$ symmetry maps one boundary to the opposite boundary, such that any state localized at only one of the boundaries automatically breaks $PT$ \cite{Hu2011Absenceof}.

It is worthwhile to point out that skin states do not necessarily have to accumulate on one boundary alone \cite{song2019nonhermitian,HoHeScSaetal2019}. For example, taking two time-reversed copies of a cBBC-breaking model immediately results in the appearance of skin states on both boundaries, which was referred to as the $\mathbb{Z}_2$ skin effect by \onlinecite{OkumaKawabataShioSato2019}. Additionally, skin states may also appear on boundaries with a codimension higher than 1, such as at corners and hinges \cite{EdKuBe2019, Ez2018, Luo2019,Liu2019}.

\subsubsection{Spectral instability} \label{subsec:spectral_instability}

As with the eigenstates, the eigenvalues of cBBC-breaking NH Hamiltonians are extremely sensitive to perturbations that connect boundaries; see the inset in Fig.~\ref{fig:abs_spec_nh_ssh_mod}(b). This sensitivity to boundary conditions can even result in drastically different qualitative features of the two spectra: Indeed, the OBC spectrum may, for instance, be gapped and topologically nontrivial, while the PBC spectrum of the same model is gapless \cite{KuEdBuBe2018}. This spectral instability can be systematically understood as a discontinuous behavior in the eigenvalue spectra of NH matrices under small random perturbations. More specifically, while adding a small perturbation with largest absolute eigenvalue $\epsilon$ to a Hermitian system at most leads to a change of the order of $\epsilon$ in the spectrum, in NH matrices changes of the order of $\epsilon^{1/N}$ may occur \cite{Krause1994}, where $N$ is the number of sites. In the thermodynamic limit ($N \rightarrow \infty$), this amounts to a change of the order of 1 for an arbitrarily small $\epsilon > 0$ representing the analytical reason for the observed fragility of eigenvalue spectra in NH systems. The tuning between boundary conditions may be interpreted as such a perturbation; see \onlinecite{HeBaRe2019} for a detailed discussion.

This spectral instability is related to the previously discussed NH skin effect: A study of the spectral instability in a cBBC-breaking model revealed that when tuning between OBCs and PBCs, one or more higher-order EPs are encountered \cite{Xiong2018}. Indeed, when the boundaries of a NH model with OBC are connected via an exponentially small perturbation proportional to the system size $N$, i.e., $\sim e^{-\alpha N}$ for some model-dependent constant $\alpha$ \cite{KuEdBuBe2018, Koch2019}, the spectrum shows crossover behavior, which can be understood from the behavior of the skin states: For a large enough coupling, which is found to be exponentially small in $N$, the skin states can tunnel through and behave like ordinary bulk states in the sense that they are evenly distributed throughout the lattice, in which case the spectrum qualitatively resembles that of the PBC case \cite{KuEdBuBe2018}. Additionally, the presence of perturbations connecting the boundaries was shown to result in unconventional behavior for the fidelity and Loschmidt echo near the higher-order EPs \cite{Longhi2019}.

Because of the extreme sensitivity of cBBC-breaking NH models to boundary conditions, it seems natural to wonder about the physical relevance of studying the eigenvalue spectra of such NH models with OBCs \cite{gong,HeBaRe2019}. However, when requiring physically motivated locality conditions on the considered perturbations, the physical properties specific to the eigenspectra of NH systems with OBCs have been shown to be robust \cite{Koch2019}. This renders the anomalous BBC observed in the eigenvalue spectra of NH systems a topologically stable and generically observable phenomenon.

\begin{figure*}[t]
  \includegraphics[width=0.9\textwidth]{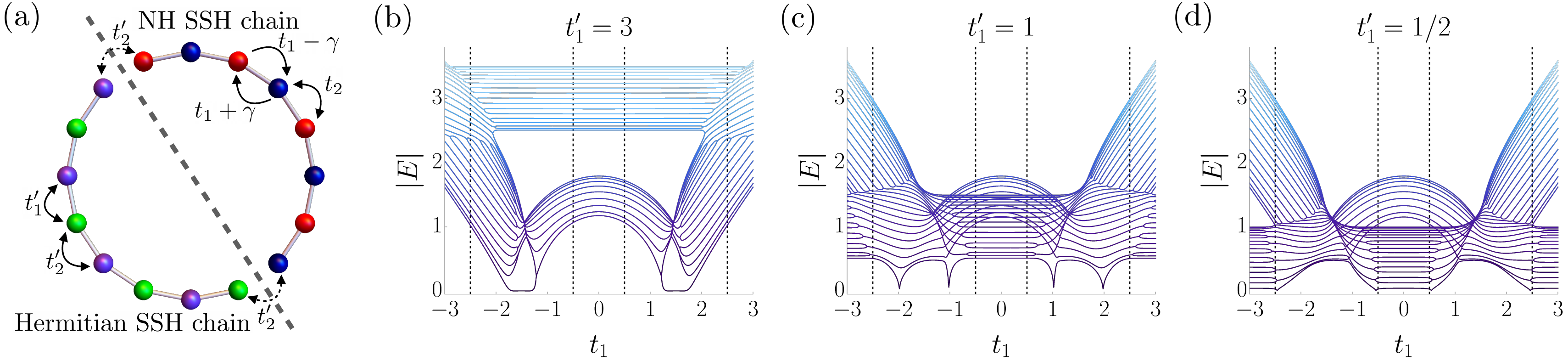}
  \caption{(a) Domain-wall geometry between a NH-SSH domain [cf. Eq.~(\ref{eq:nh_ssh_ham_real_space})] [top half circle with lattice sites in red and blue (gray and black)] and a Hermitian region [bottom half circle with sites in green and purple (light and dark gray)]. The NN hopping parameters for the NH-SSH chain are $t_1$, $t_2$, and $\gamma$ while in the Hermitian part they are $t_1'$ and $t_2'$. The two chains are coupled to each other via the hopping parameter $t_2'$. (b)--(d) Absolute value of the eigenvalues as a function of $t_1$ for $t_2 = 1$, $\gamma = 1.5$, $t_2'=0.5$, and $N = 18$ unit cells in both chains, and (b) $t_1'=3$, (c) $t_1' = 1$, and (d) $t_1'=1/2$. The spectrum in the Hermitian SSH chain is (b) gapped, (c) gapped with a smaller gap and (d) gapless. The black, dashed grid lines correspond to the gap closings in the PBC spectrum of the NH-SSH chain showing that even though the NH model is coupled to a Hermitian chain via a domain wall, the anomalous physics persists when the gap in the Hermitian chain is large enough.}
  \label{fig:domain_wall}
\end{figure*}

\subsubsection{Domain-wall geometries}\label{sec:domainwalls}

Thus far we have focused on the physics of cBBC-breaking NH models in the case of PBCs and OBCs and noticed that both the quantitative and qualitative behavior of these NH models can be extremely different in these two cases. Another interesting geometry to consider is that of \emph{domain walls}, which can lead to drastic alterations of the physics of NH models \cite{Schomerus2013Topologically, Malzard2015Topologically, Malzard2018Bulk,Deng2019}. For example, \onlinecite{Xiong2018} pointed out that if a cBBC NH model is coupled to another model that resides in a different topological phase, high-order EPs disappear rapidly from the spectrum. It has been conjectured \cite{Xiong2018,Leykam} that cBBC is generically restored in such domain-wall geometries. However, \onlinecite{KuEdBuBe2018} explicitly exemplified that upon coupling the NH-SSH model [cf. Eq.~(\ref{eq:nh_ssh_ham_real_space})] to its Hermitian, topologically trivial counterpart, cBBC may remain broken in the sense that the skin effect prevails. Indeed, the proximity to EPs persists and bulk states still locally accumulate, albeit now at the domain wall, as long as the energy gap in the Hermitian system is large enough \cite{KuEdBuBe2018}; see Fig.~\ref{fig:domain_wall}. For a sufficiently small gap (or short Hermitian domain), the skin states can tunnel through, and behavior similar to the NH model with PBC is retrieved \cite{KuEdBuBe2018,HeBaRe2019}. Changing the size of the band gap in the attached Hamiltonian in such a setup may thus be seen as an alternative way of tuning between OBCs and PBCs, while at the same time introducing new effects that cannot be observed in simple OBC geometries.

Domain-wall effects have also been studied for models that preserve cBBC. For example, by coupling two $PT$-symmetric SSH chains that are in distinct topological phases, a defect state appears at the domain wall with positive imaginary energy, thus representing a solution with a growing amplitude, while the bulk states all have zero imaginary energy \cite{Schomerus2013Topologically}. A similar setup considered by \onlinecite{YucePRA2018} confirmed that $PT$ symmetry is indeed spontaneously broken on the interface. As a consequence, the defect state dominates in the long-time limit. These predictions were experimentally confirmed in a resonator chain \cite{Poli2015Selective}, paving the way to the experimental realization of topological lasers \cite{StJean2017Lasing, Zhao2018Topological, Parto2018EdgeMode}. Additionally, it was shown that models that are topologically trivial in the Hermitian limit can host topologically protected defect states in the NH case \cite{Malzard2015Topologically, Malzard2018Bulk}. Considering domain walls in the form of defects can thus lead to new genuinely NH physical phenomena.

\subsection{Approaches to reestablishing the bulk-boundary correspondence in NH systems}
\label{sec:approaches_nh_bbc}

While the concept of a BBC in NH models with a point gap in their complex spectra has largely remained elusive, significant progress has been made on reestablishing a NH BBC in models with line gaps \cite{KuEdBuBe2018, YaWa2018, YaSoWa2018,Xiong2018}, which is the focus of our subsequent discussion. We review two main approaches in detail: (i) a construction combining information about both open and translation-invariant systems that leads to modified topological invariants akin to those in the Hermitian realm \cite{YaWa2018, YaSoWa2018}, and (ii) the biorthogonal BBC approach, which makes direct use of the properties of the OBC spectrum and relates phase transitions to delocalization transitions of biorthogonal boundary states \cite{KuEdBuBe2018}. While they are seemingly distinct we elucidate the equivalence of these two approaches, which from a different angle provide accurate predictions for generic NH systems. We also give an overview of complementary works relating to NH BBC \cite{HeBaRe2019,Lieu,Hyart,Esaki2011,LeTh2019,Borgnia,Rosenow,Imura,EdKuBe2019,Kunst2019,YaSoWa2018,Murakami,Yang2019b,song2019nonhermitian}.

\subsubsection{Non-Bloch bulk-boundary correspondence} \label{subsect:non_bloch_BBC}

A strategy for finding a generalized BBC was presented by \onlinecite{YaWa2018, YaSoWa2018} and further expanded upon by \onlinecite{Murakami}, \onlinecite{Yang2019b}, \onlinecite{Deng2019} and \onlinecite{Kawabata2020}. There a generalized BZ is constructed to include information, which in the case of cBBC is not contained in the standard Bloch bands, pertinent for the accurate definition of bulk topological invariants. The key idea in this approach is that a state with degree of freedom $j$ in unit cell $n$ of a model with OBC $\psi_{n,j}$ can be written as $\psi_{n,j} = \beta_j^n \psi_j$, where $\beta_j \equiv r_j e^{i k}$ and $\psi_j$ is the eigenvector of the Bloch Hamiltonian. Solutions for $\beta_j$ in terms of the hopping parameters and energy eigenvalues are then found by solving the eigenequations using this ansatz \cite{YaWa2018}. From these solutions, it is possible to derive the generalized BZ $C_\beta$, and to find expressions for the boundary states, as we review in the following.

The generalized BZ is found by looking at the condition for obtaining the continuum bands \cite{Murakami}. Ordering the solutions $\beta_j$ according to $|\beta_1| \leq |\beta_2| \leq \cdots \leq |\beta_{2S-1}| \leq |\beta_{2S}|$, where $S = \alpha L$ with $\alpha$ degrees of freedom and $L$ is the range of hopping, \onlinecite{Murakami} proposed that the continuum states are retrieved by demanding that
\begin{equation}
|\beta_S| = |\beta_{S+1}| = r. \label{eq:non-blochBBC_continuum_state_condition}
\end{equation}
This condition is derived by assuming that the system size $N$ is large and the energy states are densely distributed. The complex-valued trajectories of $\beta_S$ and $\beta_{S+1}$ then form the generalized BZ $C_\beta$, which in the case of Hermitian or cBBC-preserving NH Hamiltonians simply reduces to the unit circle, i.e., to the conventional one-dimensional BZ.

When $|\beta| \neq 1$, where we have dropped the label to simplify notation, the continuum states exhibit the NH skin effect: They localize to the left boundary for $|\beta| <1$ and to the right boundary for $|\beta| >1$. As mentioned, it is also possible to find models with skin states that are localized to opposite boundaries (cf. Sec.~\ref{subsubsec:skin_effect}), in which case part of the generalized BZ $C_\beta$ lies inside the unit circle, and part of it outside \cite{song2019nonhermitian}.

If the energy $E_\textrm{top}$ of possible topological boundary modes is known, it is possible to find a solution for these states by plugging $E_\textrm{top}$ into the solutions $\beta_j$ \cite{YaWa2018}. The bulk-band gap then has to close when $|\beta(E_\textrm{top})| = r$, i.e., when the topological boundary state merges with the bulk bands.

As the energy of the boundary states is not usually known, however, an alternative way to find the band-gap closing is to make use of what \onlinecite{YaWa2018} called \emph{non-Bloch topological invariants}: Replacing $e^{i k}$ with $\beta$ or, equivalently, applying a shift in the wave vector $k \rightarrow k - i \, \textrm{ln} \, r$ in the Bloch Hamiltonian $H(k)$ leads to the so-called non-Bloch Hamiltonian
$H(\beta)$ defined on the generalized BZ and allows for the computation of non-Bloch topological invariants, which correctly predict the existence of topological boundary states. Indeed, it was exemplified that a winding number derived for $H(\beta)$ on the generalized BZ correctly predicts the existence of the zero-energy end states for the model in Eq.~(\ref{eq:nh_ssh_ham_real_space}) by \onlinecite{YaWa2018}, and a variation of the model was givenby \onlinecite{Murakami}. Furthermore, \onlinecite{Yang2019b} elaborated on the geometrical interpretation of the generalized BZ, \onlinecite{Lee2019c} derived physical responses based on this picture, and \onlinecite{YaSoWa2018} introduced a non-Bloch Chern number that accurately predicts the existence of chiral edge states.

We note that the ansatz $\psi_n = \beta^n \psi$ for the wave function, which is the basis of the generalized BZ construction, can be seen as a generalization of the usual ansatz $\psi_n = e^{i k n} \psi$ in Hermitian systems, obtained by shifting the wave vector $k$ according to $k \rightarrow k - i \, \textrm{ln} \,r$. Indeed, for Hermitian and cBBC-preserving systems, $r=1$, such that $\beta = e^{i k}$ and Bloch's theorem is retrieved. In this case, the condition in Eq.~(\ref{eq:non-blochBBC_continuum_state_condition}) is trivially satisfied for all $\beta_i$, showing that this approach connects to the well-established Hermitian limit in the expected way.

\subsubsection{Biorthogonal bulk-boundary correspondence} \label{subsect:biorthogonal_BBC}

{\textit{Biorthogonal quantum mechanics.}}---An alternative approach for finding a generalized BBC was presented by \onlinecite{KuEdBuBe2018} in the form of a biorthogonal BBC, and further generalized by \onlinecite{EdKuBe2019, EdYoKuBe2019}. To discuss the biorthogonal BBC in a self-contained manner, we recall basic elements of biorthogonal quantum mechanics (QM); see \onlinecite{Brody2013} for a pedagogical review. Biorthogonal QM can be seen as a generalization of ordinary QM by allowing for the treatment of NH observables, and it reduces to ordinary QM upon restoring Hermiticity. As mentioned in Sec.~\ref{sec:intro}, a NH Hamiltonian in general has inequivalent right and left eigenvectors $\ket{\psi_R}$ and $\bra{\psi_L}$, respectively, such that its eigenvalue equations read
\begin{equation*}
H \ket{\psi_{R,i}} = E_i \ket{\psi_{R,i}}, \qquad \bra{\psi_{L,i}} H = E_i \bra{\psi_{L,i}},
\end{equation*}
where the latter expression is alternatively written as $ H^\dagger \ket{\psi_{L,i}} = E_i^* \ket{\psi_{L,i}}$. As shown in our minimal example in Sec.~\ref{sec:intro}, the left and right eigenvectors generally do not form an orthonormal set with the standard inner product; see Eq.~(\ref{eqn:toyevecs}). However, the essence of biorthogonal QM is that, away from exceptional degeneracies, the sets $\{\ket{\psi_R}\}$ and $\{\ket{\psi_L}\}$ form a useful \emph{biorthogonal basis} by demanding that
\begin{equation}
\braket{\psi_{L,i}|\psi_{R,j}} = \delta_{i,j}. \label{eq:biorthogonal_normalization}
\end{equation}
As we see later this change in normalization condition has profound implications since the left and right eigenstates can be strikingly different and may even localize at opposite boundaries of the system. An immediate and important consequence is that the energy eigenvalues of a NH Hamiltonian are given by its expectation value with respect to the right and left wave functions, i.e.,
\begin{equation}
\braket{\psi_{L,i}|H|\psi_{R,i}}  = E_i \in \mathbb{C}.\label{eq:biorthog_expec_value}
\end{equation}
Expectation values of the form of Eq.~(\ref{eq:biorthog_expec_value}) are known as \emph{biorthogonal expectation values}, and play a central role in understanding the dynamics of NH models.

{\textit{Biorthogonal BBC.}}---In the following, we discuss how the biorthogonal formalism can be used to construct a variant of the BBC that remains intact for NH systems with a line gap and reduces to cBBC in the Hermitian limit. This approach, coined biorthogonal BBC, was introduced by \onlinecite{KuEdBuBe2018}, who showed that one way to qualitatively and quantitatively understand the physics of NH models with OBC is by making use of biorthogonal QM. 

To illustrate this method, we make explicit use of the example in Eq.~(\ref{eq:nh_ssh_ham_real_space}). \onlinecite{KuEdBuBe2018} showed for the Hamiltonian in Eq.~(\ref{eq:nh_ssh_ham_real_space}) with OBCs that it is possible to write the following ansatz for the zero-energy state, which is exponentially localized and has nonzero weight on the $A$ sublattices only:
\begin{equation}
\ket{\psi_{R/L,0}} = \mathcal{N}_{R/L} \sum_{n=1}^N r_{R/L}^n c^\dagger_{A,n} \ket{0}, \label{eq:exact_sol_right_left_nh_ssh}
\end{equation}
where $\mathcal{N}_{R}$ ($\mathcal{N}_{L}$) is the normalization factor of the right (left) wave function, $n$ labels the unit cell with a total of $N$ unit cells, and $c^\dagger_{A,n} $ creates a state in the vacuum $\ket{0}$ on sublattice $A$ in unit cell $n$. The localization factors $r_R$ and $r_L$ are different
\begin{equation}
r_R = - (t_1 - \gamma)/t_2\ \neq \ r_L = - (t_1 + \gamma)/t_2\ ,\end{equation}
and hence, depending on the parameter values, the left and right states can be localized on either the same or at opposite boundaries. It is worth noting that the possibility of having the left and right states localized at opposite boundaries implies that the biorthogonal normalization condition [Eq. (\ref{eq:biorthogonal_normalization})] becomes radically different from the standard normalization condition familiar from the Hermitian realm.

To study the localization of the zero-energy states in the lattice, the biorthogonal expectation value of the projection operator $\Pi_n = \ket{e_{A,n}}\bra{e_{A,n}}+\ket{e_{B,n}}\bra{e_{B,n}}$ with $\ket{e_{\alpha,n}} \equiv c^\dagger_{\alpha,n}\ket{0}$ projected onto each unit cell $n$ is computed and leads to $\braket{\psi_{L,0}|\Pi_n|\psi_{R,0}} = \mathcal{N}_L^* \mathcal{N}_R \left(r_L^* r_R \right)^n$ for the wave functions in Eq.~(\ref{eq:exact_sol_right_left_nh_ssh}). According to this expression, the zero-energy state is thus a bulk state when $|r_L^* r_R|=1$, i.e., when it is equally localized to all unit cells, while it is exponentially localized to $n=1$ when $|r_L^* r_R|<1$ and disappears into the bulk for $|r_L^* r_R|>1$. This indeed corresponds to what we see in the band spectrum in the inset of Fig.~\ref{fig:abs_spec_nh_ssh_mod}(b) up to finite-size corrections: The bulk gap closes when $|(t_1^2 - \gamma^2)/t_2^2|=1$, while the in-gap zero-energy states exist for $|(t_1^2 - \gamma^2)/t_2^2|<1$. Note that identical results are found when considering the biorthogonal expectation value of the projection operator with respect to the zero-energy state localized on the $B$ sublattices at the end $n=N$. Thus, $|r_L^* r_R|$ determines whether boundary states exist, defining the notion of a biorthogonal BBC. By contrast, the ordinary expectation values (based on the left and right eigenstates, respectively) yield $\braket{\psi_{R,0}|\Pi_n|\psi_{R,0}} \sim \left|r_R \right|^{2n}$ and $\braket{\psi_{L,0}|\Pi_n|\psi_{L,0}} \sim \left|r_L \right|^{2n}$, respectively, for the wave function in Eq.~(\ref{eq:exact_sol_right_left_nh_ssh}). Both of these expectation values coincidentally predict gap closings in the PBC spectrum and thus fail to correctly predict the formation of zero-energy edge modes when cBBC is broken.

{\textit{Biorthogonal polarization.}}---Generalizing the insights gained from the aforementioned quantity $|r_L^* r_R|$, \onlinecite{KuEdBuBe2018} introduced the biorthogonal polarization
\begin{equation}
P = 1 - \underset{N \to \infty}{\textrm{lim}} \frac{\sum_n \braket{\psi_{L,0}| n \Pi_n|\psi_{R,0}}}{N}. \label{eq:biorthog_pol}
\end{equation}
From this expression, it is straightforward to see that $P$ equals 1 in the presence of end states, i.e., when $|r_L^* r_R|<1$ in the previous discussion, and 0 when no such states exist, i.e., when $|r_L^* r_R|>1$. $P$ jumps when the gap closes corresponding to $|r_L^* r_R|=1$. As such, the value of the biorthogonal polarization accurately predicts the presence of boundary states inside the bulk gap, and can thus be interpreted as a \emph{real-space invariant}.

We note that the condition $|r_L^* r_R|=1$ is equivalent to the merging condition [$|\beta(E_\textrm{top})| = r$] found within the non-Bloch framework \cite{YaWa2018}: Indeed, the anistropic SSH model in Eq.~(\ref{eq:nh_ssh_ham_real_space}) was also studied by \onlinecite{YaWa2018} leading to equivalent results for the topological boundary states as well as their attachment to the bulk bands.

We note that the biorthogonal polarization $P$ is equal for models that are related to each other via unitary transformations acting locally, e.g., $P_\textrm{SSH}$ for the nonreciprocal SSH model equals $P_\textrm{Lee}$ for Lee's model discussed in Sec.~\ref{subsubsec:canonical_examples} \cite{EdYoKuBe2019}.

{\textit{Generalizations.}}---As pointed out by \onlinecite{KuEdBuBe2018}, the wave-function solution in Eq.~(\ref{eq:exact_sol_right_left_nh_ssh}) can straightforwardly be generalized to a large family of lattice models with any dimension such as NH Chern insulators in two dimensions. Further generalizations to higher-order boundary states of NH models work analogously \cite{EdKuBe2019}: In each case $|r_L^* r_R|$ determines the existence of boundary states and accurately predicts the occurrence of phase transitions. It has also been verified that the definition of the biorthogonal polarization can be naturally extended to models with multiple boundary states on one boundary \cite{EdYoKuBe2019}.

We emphasize that the biorthogonal polarization defined in Eq.~(\ref{eq:biorthog_pol}) is not limited to solutions of the form given in Eq.~(\ref{eq:exact_sol_right_left_nh_ssh}) but can be computed for any boundary state in generic NH models that do not afford an exact analytical solution \cite{KuEdBuBe2018}. This makes the biorthogonal BBC a general principle for NH topological models, which recovers the cBBC where applicable.

Last, we note that while right wave functions are most naturally accessible in experiment, \onlinecite{Schomerus2019} proposed in a recent theoretical work that it is also possible to probe left wave functions as well as the biorthogonal contribution of both right and left wave functions beyond the spectral properties when measuring the response functions to external perturbations in robotic metamaterials such as the ones studied by \onlinecite{Brandenbourger2019, GhBrWeCo2019}; see Sect.~\ref{sec:exp_mechanics} for a more detailed discussion. Exploiting these possibilities it was suggested that NH topological sensors with a robust sensitivity scaling exponentially with the size of the system may be realized \cite{Budich2020}.

\subsubsection{Complementary approaches} \label{sec:unconventional_BBC_additional_approaches}

We now give an overview of complementary perspectives and approaches to BBC in NH systems reported in the recent literature.

\emph{Refinements.}---The NH BBC developed by \onlinecite{KuEdBuBe2018,YaWa2018} has been refined and corroborated by a number of recent studies.
As discussed in Sec.~\ref{subsect:biorthogonal_BBC} for the biorthogonal approach, it is beneficial to have access to exact solutions to understand the properties of NH models. As a complementary approach to obtaining such exact solutions, transfer-matrix methods were introduced in the context of NH models by \onlinecite{Kunst2019}. There one of the central results is that the determinant of the transfer matrix $T$ associated with a given NH hopping model plays a crucial role in determining whether cBBC is broken: Namely, when the transfer matrix is unimodular, i.e., $|\textrm{det}T|=1$, the PBC and OBC spectra are equivalent, and bulk states in the OBC case behave in the ordinary fashion. When $|\textrm{det}T|\neq1$, on the other hand, more interesting properties arise: The bulk spectra for PBCs and OBCs are different, while the norm of the bulk states in the OBC case is proportional to $|\textrm{det}T|^{n/2}$, with $n$ labeling the supercell, thus signaling the NH skin effect. It is possible to tune between the bulk spectra by applying a shift to the crystal momentum, i.e., $E_\textrm{PBC}(k) \rightarrow E_\textrm{OBC}(k)$ when $k \rightarrow k - (i/2)\textrm{log}(\textrm{det} T)$, where this shift in the Bloch momentum is equivalent to the one found by \onlinecite{YaWa2018} thus corroborating the generalized BZ approach. The transfer-matrix method also corroborates the findings from the biorthogonal approach: The eigenvalues of the transfer matrix for the boundary states, which correspond to the decay coefficients of their boundary states, naturally lead to the definition of a merging condition equivalent to the one found by \onlinecite{KuEdBuBe2018} ($|r_L^* r_R|=1$). Additionally, transfer matrices can be used to determine the appearance of EPs in the OBC spectrum. Indeed, when $|\textrm{det}T|\rightarrow 0, \infty$ it is possible to hop in only one direction and EPs with an order scaling with system size naturally show up in the OBC spectrum; see also the discussion in Sec.~\ref{subsubsec:skin_effect}.

The biorthogonal and non-Bloch frameworks were further expanded by \onlinecite{LeTh2019}, where a complex flux effectively interpolating between PBCs and OBCs was used \cite{Hatano1996}, which is equivalent to tuning the value of the complex part ($- \textrm{ln} \, r$)
of the complex momentum ($k - i \, \textrm{ln} \, r$) as introduced by \onlinecite{YaWa2018}. The insertion of the complex flux allows for the derivation of a condition for the existence of bulk and, more particularly, skin states akin to the one in Eq.~(\ref{eq:non-blochBBC_continuum_state_condition}).

The BBC of NH models was also studied by making use of Green's functions or, more specifically, boundary Green's functions by \onlinecite{Borgnia, Rosenow} to find topological phase diagrams. In particular, \onlinecite{Rosenow} used this machinery to study NH Dirac fermions in one dimension, and a nonzero winding number [cf. Eq.~(\ref{eqn:spectralwinding}), with $E_k$ replaced by $\textrm{det} H(k)$] is found to lead to a spatial growth of the bulk Green's function signaling a breakdown of cBBC and the occurrence of the NH skin effect. This relation between a nontrivial winding number and the appearance of skin states has indeed been elaborated upon at the level of dynamic matrices \cite{Wanjura2019} as well as at the Hamiltonian level in recent works \cite{ZhangYangFang2019, OkumaKawabataShioSato2019}; cf. Sect.~\ref{subsubsec:skin_effect}. \onlinecite{Borgnia} found edge modes by computing the in-gap zeros of the doubled boundary Green's function, where the input Hamiltonian is of the form of Eq.~(\ref{eq:doubled_hamiltonian}). There, by studying the Green's function in this framework, a classification of NH models in terms of their gaps is found, thus extending the results given by \onlinecite{Rosenow}. 

As a complementary approach, \onlinecite{Imura} proposed modified periodic boundary conditions (mPBCs) to restore BBC in NH systems. The key idea is that the mPBCs incorporate the NH skin states directly into a modified periodic model from which it is then possible to compute topological invariants that accurately predict the existence of boundary states in the case of OBCs. The mPBCs by \onlinecite{Imura} bear some similarity to the argument of imaginary flux threading by \onlinecite{LeTh2019}: The mPBCs are implemented through the inclusion of prefactors $r^L$ and $r^{-L}$ in the Hamiltonian that connects the two ends $n=1$ and $n=N$, while the flux threading essentially introduces a similar prefactor to the Hamiltonian.
While this mPBC method seems to be similar to the non-Bloch BBC introduced by \onlinecite{YaWa2018}, there is nevertheless a subtle difference: To establish the non-Bloch BBC reference is made to a system with OBC to find the relevant $\beta$ needed to compute the non-Bloch topological invariants. In the context of mPBC, by contrast, no reference to OBC is required to find the topological invariants.

\emph{Alternative perspective: Singular value spectrum.}---\onlinecite{HeBaRe2019, Herviou2019,Porras2019} proposed to infer the topological phase diagram and the existence of boundary modes by a singular value decomposition (SVD). There the role of the eigenvalues of a NH matrix is replaced by its singular values that do not exhibit the aforementioned spectral instability, and the counterpart of the eigenvectors may be directly inferred from the transformation matrices of the SVD. This allows not only for the stable computation of topological invariants, which are constructed by making use of a generalized flattened singular decomposition, but also for a generalization of the concept of the entanglement spectrum to the realm of NH models \cite{HeBaRe2019}. However, the SVD approach leads to a restoration of cBBC, even in models where cBBC is found to be broken when studying the eigenvalue spectrum. Thus, the exotic features displayed by cBBC-breaking models are not fully captured within the SVD perspective.

\emph{Symmetries.}---The influence of symmetries on BBC in NH has been widely studied \cite{Esaki2011, Lieu, Hyart,Kunst2019, Kawabata2018} (see also Sect.~\ref{subsubsec:skin_effect}), and cBBC has been shown to be preserved in a number of symmetric NH models. For example, \onlinecite{Esaki2011} showed that even though the spectrum of NH systems generically is complex, it is possible to find topological invariants from the Bloch Hamiltonian that accurately predict the existence of boundary states in the real part of the spectrum for specific lattice models, which either feature pseudo-Hermiticity [cf. Eq.~(\ref{eqn:PseudoHSym}) with $Q_-$] or time-reversal symmetric with the time-reversal operator $T_+$ [cf. Eq.~(\ref{eqn:AZ_TRS})]. In some NH systems even the TRS of type $T_+$ leads to a \emph{generalized Kramers theorem} \cite{Sato2012}. A related form of the pseudo-Hermitian symmetry $Q_-$ was investigated by \onlinecite{Hyart}, who studied a special form of NH chirality, i.e., $S H(k) S = - H^\dagger(k)$. By considering one-dimensional models in the presence of this symmetry, a hidden Chern number can be defined that determines the number of end states whose real part of the energy is zero. There the imaginary part of the energy is used as a second dimension, which offers a new perspective on the definition of topological invariants in NH models. \onlinecite{Leykam, LeeSkinEffect2016} also studied chiral symmetry and found half-integer winding numbers characterizing the EPs in the spectum. \onlinecite{Lieu} studied both chiral symmetry and $PT$ symmetry in the context of NH variations to the SSH model, and topological invariants derived from the Bloch Hamiltonian have been found in both cases. More specifically, a global invariant can be defined in the $PT$-symmetric case, while a quantized complex Berry phase exists in the case with chiral symmetry. With the more recent studies of the NH model in Eq.~(\ref{eq:nh_ssh_ham_real_space}), which is also chirally symmetric, however, we know that such a complex Berry phase cannot always be found, or at least needs to be modified by using the techniques developed in the non-Bloch setting \cite{YaWa2018, YaSoWa2018, Murakami}.

\subsubsection{Summary: A unified picture}

Having reviewed various complementary approaches to reestablishing the BBC in NH systems, we now summarize the different methods by drawing a unified picture. Whereas the main approaches introduced by \onlinecite{KuEdBuBe2018} and by \onlinecite{YaWa2018, YaSoWa2018}, respectively, have different vantage points, we stress that they lead to identical predictions in full agreement with a wide range of explicit model calculations. 

While \onlinecite{KuEdBuBe2018} took a direct cue from the properties of systems with OBCs and examined the \mbox{(de)}localization transitions of the biorthogonal wavefunctions, \onlinecite{YaWa2018, YaSoWa2018} instead augmented the Bloch Hamiltonian with information from the OBCs leading to a generalized Brillouin zone (often called non-Bloch) description that relates more directly to the familiar picture of the cBBC in terms of topological invariants. The biorthogonal approach, on the other hand, offers additional physical insights in terms of a quantized polarization and reveals the key role played by the interplay between left and right wave functions, a distinction that is inherently NH. Taken together they thus offer a comprehensive framework and physical intuition for cBBC-breaking NH models. Moreover, despite their differences in appearance, these approaches do share the emphasis on a wave-function ansatz, which were also utilized and expanded on by \onlinecite{Kunst2019, LeTh2019, Imura}. 

Several recent works have corroborated and elucidated the non-Bloch approach either by making the Bloch momentum complex \cite{Murakami, Kunst2019}, or by applying mPBCs \cite{Imura}. Complementing this perspective \onlinecite{Borgnia, Rosenow} made a direct connection to OBCs thus being conceptually more in line with biorthogonal approach, while work by \onlinecite{Kunst2019,LeTh2019} interpolated between PBC and OBC cases, and may as such be seen as a bridge between the approaches using Bloch Hamiltonians and those using OBC descriptions.

\section{Physical platforms}\label{sec:NH_applications} 
We now give an overview of experimental platforms for the observation of NH topology, reflecting the range of physical incarnations of the genuinely NH phenomena.

\subsection{Non-Hermitian wave equations: From classical mechanics to quantum walks}
Intense research in recent years has unraveled classical analog of topological phases in a variety of settings ranging from
photonics \cite{Haldane2008,Raghu2008,Ozawa2019} to electric circuits \cite{Ningyuan2015, Albert2015,Lee2018} and mechanical systems \cite{Huber2016,Kane2013}. Guided by the intuition from ideal dissipation-free scenarios, such analogs were initially established for nearly Hermitian systems. However, in all of these settings non-Hermiticity actually occurs naturally, reflecting the ubiquitous role of dissipation. Indeed, the profound conceptual advances in understanding NH topological phenomena as discussed in this review have been closely accompanied by corresponding experiments in all of the aforementioned platforms. In these classical systems, the analogy with Hamiltonian QM may manifest in a number of different ways: Some settings, including optical waveguides, directly mimic the time-dependent Schr\"odinger equation, while in photonic crystals and acoustic systems the eigenmode problem is tantamount to the Bloch problem familiar from quantum systems with a periodic potential. Similarly, in robotic mechanical metamaterials the analog of a QM Hamiltonian is directly given by an asymmetric dynamical matrix, while in electrical circuits the analogy is on the level of response functions. While we refer to the original work for details, we outline here some of the basic ideas behind the various experimental applications to make this overview more self-contained.

\subsubsection{Photonics} \label{sec:exp_photonics}

Photonics is arguably the area in which NH topology has thus far found most applications. For an in-depth account on mostly Hermitian topological photonics see the recent review given by \cite{Ozawa2019}. We highlight here a few systems with particular relevance to the genuinely NH phenomena. 

We begin with {\it photonic crystals}, in which the basic idea is to create metamaterials with spatially varying but periodic dielectric permittivity $\epsilon_{ij}(\mathbf x)$ and magnetic permeability $\mu_{ij}(\mathbf x)$ \cite{Joannopoulos2011}. In this setting the electrodynamic eigenmodes of Maxwells equations are subject to Bloch's theorem in a manner similar to how it applies to electrons in crystalline solids. Inspired by the seminal theoretical proposal for photonic analogs of quantum Hall states due to \onlinecite{Haldane2008,Raghu2008} and subsequent refinements by \onlinecite{Wang2008}, classical analogs of topological states have been realized in gyromagnetic photonic crystals, which explicitly break time-reversal symmetry \cite{Wang2008,Lu2013}. In these systems gain and loss is ubiquitous and NH topological phenomena have been experimentally realized including a spectacular observation of Fermi arcs connecting EPs \cite{Zhou2018} as theoretically described in Sec.~\ref{sec:nodalphases}, as well as a demonstration of one-sided invisibility in $PT$-symmetric metamaterials \cite{Feng2013} predicted to occur in $PT$-symmetric materials operating at an EP \cite{Kulishov2005, Longhi2011, Lin2011, Jones2012}, which had also been shown in a scattering experiment \cite{Regensburger2012}; cf. Sect.~\ref{subsec:quantum_many_body_systems}. Recent theoretical work has suggested that the Maxwell waves existing on the interfaces separating lossless media with different signs in the permittivity and permeability have topological properties that are related to the properties of a NH helicity operator \cite{Bliokh2019} thus further highlighting the NH character of photonic crystals.

Photonic crystals belong to the larger experimental platform of \emph{optical microresonators}, also known as microcavities \cite{Vahala2003}. The performance of such resonators is captured by the $Q$ factor, which is proportional to the lifetime of a photon inside the cavity and is strongly dependent on the properties of the interface between the cavity volume and the outside. A coupled-microresonators setup with auxiliary resonators with gain and loss has been proposed to realize the Hatano-Nelson model \cite{Longhi2015}, whereas active steering of topological light has been demonstrated in two-dimensional lattices of microresonators with reconfigurable gain and loss domains \cite{Zhao2019}. One prominent example of optical microcavities with high $Q$ factors is that of whispering-gallery-mode resonators (WGMRs) \cite{Lefevre-Seguin1997, Gorodetsky1996, Vernooy1998, Knight1995,Vernooy1998a}, which derive their name from their acoustical counterpart: Electromagnetic waves are captured in the cavity because of total internal reflection.

Recently NH experimental setups of such WGMRs were proposed and observed to exhibit unidirectional lasing \cite{asymhop1,Peng2014}, single-mode lasing in $PT$-symmetric setups \cite{Feng2014, Hodaei2014}, and enhanced sensitivity against perturbations in cavities operating at second-order EPs \cite{asymhop2} due to the nonanalytic behavior of their dispersion \cite{Wiersig}. Similar behavior has also been demonstrated in higher-order EPs realized in an arrangement of coupled microring resonators \cite{Hodaei2017}.

Optical resonators operating at microwave frequencies are known as \emph{microwave cavities}, and recently the dynamical encircling of second-order EPs was studied in such a setup, revealing experimental signatures of mode switching \cite{Doppler2016} (as we saw in the minimal example in Sec.~\ref{sec:intro}). Additionally, open microwave disks form an ideal platform to study the quantum-classical correspondence in open systems, and experiments on such models demonstrate that classical quantities can describe their quantum properties and vice versa \cite{Potzuweit2012, Lu1999, Pance2000, Barkhofen2013}.

{\it Coupled waveguides} provide another versatile setting that, instead of simulating static properties, directly emulates the time evolution of tailor-made lattice models \cite{Davis1996,Longhi2009,Christodoulides2003}. The waveguides are routinely inscribed in silica glass using femtosecond lasers and have the additional appealing feature that they operate well at optical frequencies visible to the human eye \cite{Szameit2010}. Here Maxwell's equations describing the propagation of light in the $z$ direction amount to the paraxial equation
\begin{equation*}
i\partial_z\mathcal E=\left ( -\frac{1}{2k_0}(\partial_x^2+\partial_y^2)-\frac{k_0\Delta n(x,y)}{n_0}\right )\mathcal E,
\end{equation*}
which is formally identical to the two-dimensional Schr\"odinger equation with the propagation direction $z$ playing the role of time $t$, and the wave function $\mathcal E$ is the envelope of the electric field polarized along $\mathbf e$ such that $\mathbf E(x,y,z)= \mathcal E(x,y,z)e^{i(k_0z-\omega t)}\mathbf e$ is assumed to be slowly varying in the sense that $|\nabla \mathcal E|\ll |k_0\mathcal E|$, with $k_0\approx k_z\gg k_{x,y}$. The effective potential $V(x,y) \propto \Delta n(x,y)$ can be tailor-made by carving waveguides using accurate femtosecond lasers, which create a strong spatial dependence of the local refractive index $\Delta n(x,y)$. In the limit of spatially sharp carving and weak evanescent coupling between the waveguides, this system is accurately modeled by a tight-binding Hamiltonian whose hopping parameters depend on the setup and the wavelength $\lambda$ of the light. This setup has been harnessed to emulate a large number of Hermitian topological phases \cite{Rechtsman2013,Noh2015, Noh2018, ElHassan2019} and, including staggered patterns of gain and loss in the wires, the time evolution of effectively NH models has also been successfully simulated. This includes the experimental realization of exceptional rings \cite{Cerjan2019} (cf. Sec.~\ref{sec:nodalphases}), defect states in NH-SSH chains \cite{NHexp} (cf. Sec.~\ref{sec:domainwalls}), topological phase transitions \cite{Zeuner2015}, and $PT$-symmetric flatbands \cite{Biesenthal2019}, whereas a study of the stability of corner states against gain and loss has also been proposed \cite{Ozdemir2019a}. Here it is worth noting that passive systems with only staggered loss, such as that from waveguides of alternating quality, is sufficient to generate such phases: Although the energies are confined to the lower complex half plane, a global shift can make the system effectively $PT$ symmetric in a description, where the less lossy waveguides thus effectively experience gain \cite{Guo2009,Ornigotti2014,NHexp, Feng2013, Kremer2019}. Furthermore, a truly $PT$-symmetric system has been realized by making use of \emph{optical fibers} by \onlinecite{Regensburger2012}, where the use of optical amplifiers and modulators allows for the realization of a $PT$-symmetric structure in the temporal domain.

\subsubsection{Mechanical systems}\label{sec:exp_mechanics}

Mechanical systems represent another experimental medium with which NH phases can be realized. One such system is provided by \emph{mechanical metamaterials} [see \onlinecite{Huber2016, Bertoldi2017} for recent reviews], which can be described as networks consisting of masses that are connected via springs of rigid beams and are governed by Newton's equations. Newton's equations of motion for a system of coupled oscillators $\ddot{x} = - D_{ij} x_j + A_{ij} \dot{x}$, with $x_i$ the oscillators, $A$ describing the nondissipative coupling between position and velocity, and $D$ the \emph{dynamical} matrix capturing the forces between oscillators, can be recast into the following Hermitian eigenvalue problem:
\begin{equation*}
i \partial_t \begin{pmatrix}
\sqrt{D}^T x \\
i \dot{x}
\end{pmatrix} = \begin{pmatrix}
0 & \sqrt{D}^T \\
\sqrt{D} & i A
\end{pmatrix}\begin{pmatrix}
\sqrt{D}^T x \\
i \dot{x}
\end{pmatrix},
\end{equation*}
as detailed by \onlinecite{Kane2013, Roman2015, Huber2016}. Drawing from a formal correspondence between Newton's second law and the Schr\"{o}dinger equation, it is possible to realize topological phases featuring phononic boundary states in these setups. Indeed, topological phononic modes, which were classified by \onlinecite{Susstrunk2016}, have been reported to appear at the boundaries of isostatic lattices build with springs \cite{Kane2013}, at the boundaries in models consisting of rotors and rigid beams \cite{Chen2014}, at dislocations in kagome lattices consisting of rigid plates \cite{Paulose2015}, and as helical boundary states in a setup consisting of pendula \cite{Roman2015}. When the masses are replaced by gyroscopes, one obtains a so-called gyroscopic metamaterial, which has been shown to host acoustic boundary waves analogous to the edge states of the quantum Hall effect \cite{Nash2015, Wang2015}.

Inspired by these results and the connection between the dynamical matrix and the Hamiltonian description in these setups, one can conceive of NH phononic phases: Starting from a generic NH Hamiltonian matrix with off-diagonal elements $Q$ and $\tilde{Q}$, the dynamical matrix is defined as $D = Q \tilde{Q}$ \cite{GhBrWeCo2019}. This way of writing the dynamical matrix is in close analogy to the method presented by \onlinecite{Kane2013}, who study isostatic lattices, which are mechanically critical in the sense that they are near collapse: The dynamical matrix associated with the lattice is written as $D = Q Q^T$, such that by taking the `square root' one obtains the associated Hamiltonian matrix, which has $Q$ and $Q^T$ as its off-diagonal elements. Such a dynamical matrix for NH Hamiltonians ($D = Q \tilde{Q}$), which is asymmetric ($D \neq D^T$), has been experimentally realized in \emph{robotic metamaterials} \cite{Brandenbourger2019}, which combine robotics and active materials through building lattices consisting of mechanical rotors, control systems, and springs. In such setups, the NH skin effect has been observed in a nonreciprocal realization \cite{Brandenbourger2019} as well as in a model similar to the anisotropic SSH chain described in Sec.~\ref{subsubsec:canonical_examples} \cite{GhBrWeCo2019}. Both experiments thus probe the right eigenstates of the model that they investigate. In a recent work, \onlinecite{Schomerus2019} showed by making use of response theory that it is also possible to probe the left eigenstates in these setups: Whereas right wave functions specify the spatial distribution of the response of the setup to an external excitation, the information on the strength of this response with respect to where the perturbation is located is captured by the left wave functions. When considering the overall response, which includes contributions from both the right and left wave functions, \onlinecite{Schomerus2019} showed that the NH skin effect of the zero mode is related to a phase transition at which the sensitivity to perturbations becomes critical in the sense that it diverges. The inherent biorthogonality of these systems thus leaves experimental signatures beyond the characteristic energy spectra. These two experiments by \onlinecite{Brandenbourger2019, GhBrWeCo2019} also prompted the study of the NH skin effect in elastic lattices with nonlocal feedback interactions. \onlinecite{Rosa2020} found that nonlocal control allows for bulk waves to localize at different boundaries, such that a judicious choice of interactions can result in corner localization, as illustrated in two-dimensional models. \onlinecite{Scheibner2019} showed that an antisymmetric dynamical matrix $D = - D^T$ can be realized in mechanical metamaterials with \emph{odd elasticity}, which occurs due to non-energy-conserving microscopic interactions in active media. The odd elasticity is predicted to facilitate the onset of exceptional points for an overdamped lattice as well as to sustain an elastic engine cycle for an overdamped wave \cite{Scheibner2019}, to allow for the appearance of bulk elastic waves at the boundaries of one- and two-dimensional metamaterials \cite{Zhou2019}, and to host a topological phase transition mediated by the annihilation of exceptional rings in active as well as gyroscopic metamaterials with gain and loss \cite{Scheibner2020}. In addition, a recent realization of a NH phase in mechanical metamaterials was reported on by \onlinecite{Yoshida2019}, who proposed that exceptional rings appear in mechanical metamaterials with friction.

A notion of \emph{phononic} or \emph{acoustic materials} \cite{Kushwaha1993} and \emph{metamaterials} beyond the previously outlined dynamical matrix formalism outlined above also exists and may come in many forms.
Such systems have been shown to host phononic edge states in microtubules \cite{Prodan2009}, quantum-spin-Hall edge states in the form of elastic waves \cite{Mousavi2015}, and surface acoustic waves with negative refraction index on the surfaces of a phononic version of a Weyl semimetal \cite{He2018}.
Acoustic waves may also propagate through fluids, and a setup consisting of rotating fluids arranged in a crystal was predicted to realize the chiral edge states of the quantum Hall effect \cite{Yang2015}. This experimental platform can be used to realize NH phases through the judicious implementation of gain and loss. Indeed, \onlinecite{Shi2016} realized a $PT$-symmetric model where gain is implemented via coherent acoustic sources in which they acquire full control of the EP and the accompanying unidirectional transparancy. A $PT$-symmetric acoustic metamaterial was also realized by \onlinecite{Auregan2017} in an airflow duct with gain and loss implemented through the scattering of acoustic waves of diaphragms. Similarly, \onlinecite{Rivet2018} showed that acoustic waves with constant pressure can exist in acoustic waveguides with gain and loss, while \onlinecite{Zhu2018} realized an EP in a lossy acoustic system and demonstrated unidirectional propagation. Additional theoretical proposals have been made for the realization of $PT$-symmetric second-order topological phases in acoustic metamaterials with gain and loss \cite{RosendoLopez2019, Zhang2019a}, and invisible acoustic sensors with $PT$ symmetry \cite{FleurySounaisAlu2015}.

\subsubsection{Electric circuits}
Electric circuits provide another classical platform for the realization of NH topology \cite{Ningyuan2015, Albert2015}. There, instead of properties of a Hamiltonian, one directly studies response functions, where capacitors and inductors act as Hermitian elements and resistors as well as amplifiers are anti-Hermitian. As a specific example, a current depending on frequency $\omega$ flowing through a node $i$ is governed by
\begin{equation*}
I_i (\omega)= \sum_j Y_{ij}(\omega) V_j(\omega),
\end{equation*}
where $I_i(\omega)$ and $V_i(\omega)$ are the input current and potential at node $i$, respectively, and $Y_{ij}(\omega)$ is the admittance matrix or, equivalently, the inverse impedance matrix $[Z^{-1}(\omega)]_{ij}$. Specifically, $Y_{ij}(\omega)$, with $i \neq j$, is the admittance between nodes $i$ and $j$, and $Y_{ii}(\omega)$ is the admittance between node $i$ and the ground \cite{Ningyuan2015}. This relation can be derived by making use of current conservation, i.e., the total input current needs to equal the total output current, which amounts to Kirchhoff's circuit laws.

The periodicity of the electric circuit structures allows for the use of Bloch's theorem to find wave functions, while the band structure of the circuits corresponds to the eigenvalues of the admittance $Y_{ij}(\omega)$ up to a prefactor. As such, one can interpret the admittance matrix $Y_{ij} (\omega)$ as a Hamiltonian matrix. Through the arrangement of capacitors, inductors, and other electronic building blocks available in this toolbox, it is thus possible to design circuits that mimic the physics of topologically nontrivial models. This idea was introduced by \onlinecite{Ningyuan2015} and has been used to build topological circuits whose band structures, i.e., admittance eigenvalues, also realize the band topology of the Hofstadter model \cite{Ningyuan2015, Albert2015} in the M\"{o}bius strip configuration \cite{Ningyuan2015}. More recently the SSH chain and a two-dimensional extension thereof as well as a Weyl semimetal spectrum were reported on by \onlinecite{Lee2018}, whereas corner states were realized in two-dimensional setups by \onlinecite{Imhof2018}.

These realizations of Hermitian topological phases in electric circuits have paved the way to the fabrication of NH versions thereof. Indeed, by making use of resistors and amplifiers, the NH-SSH model in Eq.~(\ref{eq:nh_ssh_ham_real_space}) was realized recently by \onlinecite{HeHoImAbKiMoLeSzGrTh2019}, who corroborated the theoretical predictions. The NH skin effect was subsequently also measured by \onlinecite{HoHeScSaetal2019}. Additional proposals have been put forward for the realization of NH honeycomb lattices with PBCs \cite{Luo2018}, NH Chern insulators \cite{Hofmann2019,Ezawa2019b}, higher-order topological models with NH skin states localized to lower-dimensional boundaries \cite{Ez2018, Ezawa2019c}, a quantum-walk simulation \cite{Ezawa2019} (see Sec.~\ref{sec:classical_hybrid_quantum_walks}), the realization of three-dimensional Seifert surfaces in four-dimensional circuit setups \cite{Li2019}, as well as the implementation of a pseudomagnetic field to probe exceptional Landau levels in NH Dirac and Weyl systems \cite{Zhang2019b}.

\subsubsection{Quantum walks} \label{sec:classical_hybrid_quantum_walks}

Quantum walks, which represent a conceptual framework rather than being limited to an experimental platform, provide another means to simulate and probe NH topological phases. Quantum walks can be seen as the quantum version of classical random walks, where the ``coin flip," which introduces the classical randomness by determining the trajectory of a particle, is replaced by a coin operator acting on the internal degrees of freedom of a particle, also known as the ``walker." The concept of the quantum walk was introduced by \onlinecite{Aharanov1993}, and quantum walks have been realized in several experimental platforms, such as trapped atoms \cite{Karski2009}, trapped ions \cite{Zahringer2010, Schmitz2009}, optical fiber networks \cite{Schreiber2010, Broome2010}, and nuclear-magnetic resonances \cite{Ryan2005}.

The dynamics of a quantum walk is captured by a Floquet operator $U$, which depends on the coin operator and is related to a time-independent effective Hamiltonian $H_\textrm{eff}$ via $U = \textrm{exp}(-i H_\textrm{eff})$. Through a suitable choice of $U$, the effective Hamiltonian $H_\textrm{eff}$ can be made \emph{topologically nontrivial}, resulting in the appearance of topological phases in quantum walks as predicted in theory \cite{Kitagawa2010, Asboth2012} and as shown experimentally in discrete-time quantum walks \cite{Kitagawa2012, Cardano2016, Barkhofen2017, Ramasesh2017, Flurin2017} [see \onlinecite{Wu2019} for a recent review], where the Floquet operator $U$ is applied to the walker at discrete time steps.

By instead considering a \emph{nonunitary} Floquet operator $U$, the effective Hamiltonian $H_\textrm{eff}$ of the model is NH, and it is thus possible to study NH phases. This idea was introduced by \onlinecite{Rudner2009} for a NH-SSH model with loss on every other site, thus realizing a passive version of a $PT$-symmetric SSH chain, where it is shown that the average displacement of the particle is quantized and associated with a topological invariant. Experiments on such nonunitary quantum walks reveal the existence of topological edge states at domain walls in a $PT$-symmetric SSH chain in an optical setup with balanced gain and loss \cite{Xiao2017}, as predicted in theory \cite{Mochizuki2016}. \onlinecite{Zhan2017} detected topological invariants, \onlinecite{Wang2019} studied dynamic quantum phase transitions in a $PT$-symmetric system, \onlinecite{Wang2019a} observed skyrmions in a $PT$-symmetric nonunitary quantum walk, and \onlinecite{Longhi2019b} predicted the appearance of the NH skin effect and a symmetry-breaking phase transition in a $PT$-symmetric discrete-time nonunitary quantum walk. Models with anisotropic hoppings have also been realized in a discrete-time nonunitary quantum-walk setup, where the NH skin effect has been explicitly detected \cite{XiDeWaZhWaYiXu2019,Weidemann2020}.

\subsection{Quantum many-body systems} \label{subsec:quantum_many_body_systems}
While most early applications of NH topology were based on classical physics and single-particle quantum mechanics, non-Hermiticity also plays an important role in genuinely quantum mechanical many-particle systems. Indeed, the study of NH Hamiltonians in this context has a long history of applications, such as in nuclear and atomic physics \cite{Rotter2009,Majorana1931,Majorana1931b,Fano,BreitWigner,FESHBACH1958357,FeshbachRev}. More recently, the relevance of these Hamiltonians to topological phases has been investigated in several quantum many-body platforms as outlined next.

\subsubsection{Open systems}
{\it Quantum master equations}.---The most natural source of non-Hermiticity in quantum many-body systems is the quantum dissipation induced by coupling the system to its environment. A realm of direct relevance involves quantum optical setups and ultracold atomic gases, where experiments are often carried out in the regime of a weak coupling to a Markovian reservoir represented by the continuum of surrounding electromagnetic field modes. 
In such situations, the relevant equation of motion for the reduced density matrix $\rho$ of the open system is the Lindblad master equation \cite{Lindblad1976} 
\begin{equation}
\partial_t \rho=i [\rho,H]+\sum_n \left( L_n \rho L_n^\dagger-\frac 1 2 \{L_n^\dagger L_n,\rho \}\right), \label{lindblad}
\end{equation}
where the jump operators $L_n$ account for the coupling to the environment. Focusing mostly on the case of pure dissipation ($H=0$), the dissipative preparation of topological states within the full Lindblad setting has been investigated \cite{TopDis,TopDis2,Budich2015,Tonielli2019,Goldstein2019}. However, owing to the complexity of the Lindblad master equation, a different approach is desirable for obtaining an intuitive understanding of the interplay between coherent quantum dynamics, dissipation, and topology in complex quantum many-body systems. To this end, one useful approach is to note that the Lindblad equation can conveniently be written as $\partial_t \rho=i (\rho H_{\rm eff}^\dagger-H_{\rm eff}\rho)+\sum_n L_n \rho L_n^\dagger , 
$ where the effective NH Hamiltonian 
\begin{equation}
H_{\rm eff}=H-\frac i 2 \sum_n L_n^\dagger L_n \label{heff}
\end{equation} 
describes the dynamics at short times \cite{Carmichael2014}. At longer times the so-called jump (or recycling) term $\sum_n L_n \rho L_n^\dagger$ accounting for the actual occurrence of quantum jumps can typically no longer be neglected. In the general situation this thus leads to decoherence (and hence mixed states), while the effective non-Hermitian description is by construction in terms of less general pure states. Nevertheless, the relevance of NH Hamiltonians for Lindblad systems reaches far beyond the obvious realm at short lifetimes: It is easy to construct intriguing examples where the steady state of the Lindblad equation is identical to the $t\rightarrow \infty$ state resulting from the nonunitary time evolution of an effective NH Hamiltonian. A simple and constructive way of achieving this is to reverse engineer models using the condition $L_n \lvert\psi\rangle =0$ \cite{TopDis}, which in effect can target for example the ground state $\lvert \psi\rangle$ of a model Hamiltonian with a suitable choice of the Lindblad jump operators. This approach is particularly well suited for preparing topological phases that quite generically have parent Hamiltonians composed of noncommuting terms that can nevertheless be simultaneously minimized. This may serve as an efficient way of harnessing dissipation and the intuition from NH Hamiltonians to realize essentially Hermitian topological phases. It is also worth noting that the effective Hamiltonian (\ref{heff}) has eigenvalues in the lower complex half-plane ${\rm Im}[E]\leq 0$. This highlights the fact that the Lindblad equation, even in the regime accurately captured by Eq.~(\ref{heff}), imposes a fundamental constraint on eligible NH Hamiltonians relative to the fully generic case; see Section \ref{phys_constraints}.

For Gaussian systems described by a Lindblad equation that is quadratic in the field operators, there is another way of systematically deriving an effective NH description in terms of a damping matrix $H_D$ \cite{Prosen2010,Eisert2010}. Complementary with the previously mentioned $H_{\rm eff}$, the NH matrix $H_D$ governs how deviations from the steady state are damped out, thus describing the long-time limit of the Lindblad equation. These two effective NH matrices have been shown to generally differ in their topological properties \cite{LindbladSkin}. In the context of Gaussian Lindbladians, genuinely NH phenomena have recently been discovered \cite{LindbladSkin,Lieu2019, Hatano2019}. A salient example along these lines is that the phenomenology of the non-Hermitian skin effect carries over, mutatis mutandis, to the more fundamental Lindblad setting \cite{LindbladSkin} where it had previously been overlooked. Moreover, exceptional points also appear naturally within the Lindblad master equation framework \cite{Hatano2019}, and certain classes of quadratic Lindblad operators admit a classification analogous to that of NH Hamiltonians \cite{Lieu2019}.

{\it Material junctions} in quantum transport setups provide another generic and conceptually clear electronic setting for realizing NH topological phases; see \onlinecite{Bergholtz2019} for a detailed discussion. In fact, the well-established theory of quantum transport that has been used and experimentally tested over decades of intense research is entirely based on NH physics; see \onlinecite{Datta2005}. A more recent development is essentially the perspective that these problems can be recast in the systematic context of NH topology, which has already inspired suggestions for novel phenomena in experimentally accessible solid-state setups.
We now consider such a setup, where one side of the junction is considered to be a thermal reservoir (lead), which induces a self-energy on the surface of the system, thus leading to the effective NH system Hamiltonian
\begin{align}
H_{\textrm{NH}}=H+\Sigma_L^r(\omega),
\label{eqn:generalHNH}
\end{align}
where $H$ is the Hermitian Hamiltonian of the isolated system and $\Sigma_{L}^r(\omega)$ denotes the retarded self-energy at energies $\omega$ close to the chemical potential reflecting the coupling to the lead. Owing to causality all eigenvalues of $\Sigma_{L}^r(\omega)$ reside in the lower half plane ${\rm Im}[E]\leq 0$. Since $\Sigma_{L}^r(\omega)$ is generically non-Hermitian and matrix valued, it can have drastic implications for the topology of the interface states. This has been investigated in the context of superconducting junctions featuring EPs \cite{Pikulin2012,Pikulin2013,Avila,San-Jose2016} as well as in interfaces between topological insulators coupled to ferromagnetic leads \cite{Gilbert,Chen2018,Bergholtz2019}. In the latter case, the Hall conductance in the gapped phase loses its quantization \cite{Gilbert,Chen2018} thus signaling a breakdown of the topological nature of the system that is well known from the Hermitian limit. However, the non-Hermiticity of this setup can also promote the topological properties: While the ferromagnet breaks time-reversal symmetry, one would expect it to generally open a gap in the surface theory. As shown by \onlinecite{Bergholtz2019}, there is a critical angle of the magnetization beyond which the dissipation overcomes the gap, thus promoting the symmetry-protected surface topology to a nodal NH topological phase with EPs and NH Fermi arcs that does not rely on any symmetry.

Photonic and hybrid systems also feature NH topology in the quantum regime. An example of this is the concept of topological lasers \cite{Bahari2017, StJean2017Lasing, Zhao2018Topological, Parto2018EdgeMode,Bandres2018,Harari2018, Longhi2018, Longhi2018a}. Lasers fundamentally depend on gain and the basic idea of topological lasers thus includes ingredients of topology, quantum mechanics, and non-Hermiticity. 

NH topology may also appear in less obvious ways, as exemplified in the bosonic Bogoliubov-de-Gennes (BdG) problem, which occurs naturally in various settings ranging from photons under parametric driving \cite{McDonald2018} to exciton polariton systems \cite{Bardyn2016} and cold atomic gases \cite{Barnett2013}. Although superficially identical to the fermionic BdG problem well known from the theory of superconductivity, the transformation needed to diagonalize the BdG Hamiltonian for bosons is paraunitary rather than unitary and the corresponding spectra are not generally real. Indeed, parametric instabilities corresponding to complex eigenvalues are known to occur in several experimentally relevant settings \cite{Barnett2013,Shi2017,Peano2016,Peano,Galilo2015}. As such, these provide a distinct raison d'\^etre for NH classification schemes, as observed by \onlinecite{LieuSym}. We note that a generic mapping between parametrically driven Hermitian bosonic models and non-Hermitian Hamiltonians beyond the BdG formalism also exists and can be used to realize NH topologically nontrivial models in Hermitian bosonic setups \cite{Wang2019b}.

Shaken cold atoms in optical lattices provide yet another platform for topological physics \cite{Eckardt2017} and atomic losses can in principle trigger the NH skin effect \cite{Li2019b}.

\subsubsection{Emergent dissipation in closed systems}
At a global level, a closed quantum mechanical system undergoing unitary time evolution does not feature dissipation. However, local observables in interacting quantum many-body systems obey nonlinear equations of motion, thus effectively leading to dissipative dynamics. In this context, it has been proposed that dissipation in the form of emergent non-Hermiticity can have a profound impact on the low-energy description of interacting and disordered quantum matter \cite{koziifu,Yoshida2018,Zyuzin2018,Michishita2020}. Phenomenologically, this scenario is reminiscent of the concept of eigenstate thermalization \cite{Deutsch1991,Srednicki1994}, a generic feature of nonintegrable quantum systems with a large number of degrees of freedom, where the system acts as its own thermal bath for local observables. In the present context, quasiparticles with a given momentum scatter off each other or at impurities and thereby acquire a finite lifetime. The corresponding self-energy is non-Hermitian and, when sufficiently generic, one may thus for example expect it to feature exceptional degeneracies and their concomitant phenomenology as discussed in Sec.~\ref{sec:two_band_NH}. 

Along these lines, suggestions about emergent topological NH phenomena have been put forward in heavy fermion systems, which are natural due to the extreme renormalization of the bare electron properties \cite{Yoshida2018}, in nodal semimetals, which, according to the general discussion in Sec.~\ref{sec:nh_metals}, provide an ideal setting for NH nodal phases \cite{Zyuzin2019,Moors2019,Zyuzin2018,Kimura2019,Yoshida2019a}, in strongly correlated Kondo materials \cite{Michishita2019}, and for magnons (the spin-wave excitations of quantum magnets), which provide another natural platform for NH topology as explored by \onlinecite{Mcclarty2019}. Bosonic BdG Hamiltonians also occur in the context of magnons, which provides an alternative way of arriving at NH phenomenology, such as in ferromagnetic materials \cite{Shindou2013}, along the lines previously discussed in the context of open systems.

Related ideas of emergent EPs were also put forward early on in the context of nodal-line semimetals in the presence of an external magnetic field \cite{Molina2018} and radiated by circularly polarized light \cite{Gonzalez2017}. Furthermore, the interplay between non-Hermiticity and superconductivity at the level of toy models has been investigated \cite{Ghatak2018}. Finally, we note that even when starting from entirely Hermitian systems, physical insights can be gained by formally extending a given model into the NH realm, as has been shown for Majorana wires \cite{Mandal2015} and interacting spin systems \cite{Luitz2019}.

\section{Concluding remarks}\label{sec:conclusion}
To summarize, bringing together insights from recent literature, in this review we have discussed how relinquishing the assumption of Hermiticity qualitatively modifies and enriches the notion of topological band structures. Both novel NH topological phases and fundamental changes to the bulk-boundary correspondence have been shown to be intimately related to the occurrence of exceptional degeneracies, a property unique to the complex spectra of NH matrices. These insights demonstrate that effective NH Hamiltonian approaches can, despite their appealing conceptual simplicity, describe intriguing topological phenomena relating to the presence of dissipation in both classical and quantum systems. This is in line with earlier findings in the fully microscopic context of quantum master equations that dissipation may be harnessed for the formation of ordered states of matter \cite{Diehl2008,Verstraete2009,TopDis} and is thus better than its destructive reputation suggests. Despite the impressive recent progress, many open questions remain in the rapidly evolving field of NH topological matter. We close our discussion by pointing out a few possible future perspectives.

Owing to the broad variety of experimental platforms for NH topological systems (see Sec.~\ref{sec:NH_applications}), a natural quest is to identify and experimentally implement potential technological applications of topological robustness and quantization in dissipative systems. As a promising step in this direction, the analytical properties of exceptional degeneracies have been reported to enhance the sensitivity of a particle detector \cite{asymhop2, Hodaei2017}. Even though a direct gain in sensing precision from the square-root dispersion around an EP has been challenged \cite{Langbein2018,Wang2020}, the general possibility of parametrically enhanced sensing due to the vicinity of EPs was reported by \onlinecite{Zhang2019c}. Recently, circumventing these issues as well as the necessity of fine-tuning to an EP, the aforementioned sensitivity to boundary conditions of NH topological systems has been harnessed for proposing a novel class of sensing devices dubbed non-Hermitian topological sensors \cite{Budich2020}. Providing another path toward new technology, topological lasers based on robust NH boundary and interface states have been discovered \cite{Bahari2017, Bandres2018,Harari2018,StJean2017Lasing, Zhao2018Topological, Parto2018EdgeMode}.

The robust quantization of response properties, most prominently exemplified by the celebrated quantum Hall effect, is a salient feature of topological phases \cite{HasanKane}. In contrast, the immediate NH generalization of a quantum Hall setting may lead to the loss of a quantized conductance \cite{Gilbert,Chen2018}. However, numerous examples of topological invariants entailing quantized observables in dissipative systems have been found \cite{Rudner2009,Hockendorf2019,Tonielli2019,Hockendorf2020,Silberstein2020}, starting with the pioneering work by \onlinecite{Rudner2009}, who connected the quantized expected displacement of a quantum walker subjected to loss to a NH winding number. Despite this progress, a general answer to the question as to what extent NH topological invariants lead to robustly quantized observables largely remains elusive, and thus represents an interesting subject of future research.

While the NH description of classical systems is satisfactorily understood within a single-particle or wave picture, the conceptually more complex case of open quantum many-body systems effectively described by a NH Hamiltonian is still far from a conclusive description. A few natural open questions in this context include the following: $(i)$ The precise relation between different levels of description, ranging from exact Liouvillian quantum dynamics to effective NH Hamiltonians, particular in the context of topological properties. $(ii)$ The presence of new topological phases beyond the independent particle picture: While intriguing NH effects in interacting systems have been reported \cite{Roncaglia2010,Yoshida2019b,Carlstrom2019,Lee2019a,Matsumoto2019,Luitz2019,Mu2019,Hanai2020,Liu2020,Shackleton2020,Yang2020}, the exploration of qualitatively new fractional topological phases that may be seen as genuinely NH counterparts to fractional quantum Hall states or spin liquids familiar from strongly correlated Hermitian systems is still largely uncharted territory.

\acknowledgements

We thank Johan Carlstr\"om, Vatsal Dwivedi, Elisabet Edvardsson, Lo\"{i}c Herviou, Kohei Kawabata, Rebekka Koch, Ching Hua Lee, Simon Lieu, David Luitz, Francesco Piazza, Stefano Longhi, Bernd Rosenow, Henning Schomerus, Marcus St\aa lhammar, Kang Yang, and Tsuneya Yoshida for discussions. E.J.B. and F.K.K were supported by the Swedish Research Council (VR) and the Wallenberg Academy Fellows program of the Knut and Alice Wallenberg Foundation. F.K.K. was also supported by the Max Planck Institute of Quantum Optics (MPQ) and Max-Planck-Harvard Research Center for Quantum Optics (MPHQ). J.C.B. acknowledges financial support from the German Research Foundation (DFG) through the Collaborative Research Centre SFB 1143 (Project No. 247310070) and the W\"urzburg-Dresden Cluster of Excellence on Complexity and Topology in Quantum Matter ct.qmat (EXC 2147, ProjectNo. 39085490).

E. J. B., J. C. B., and F. K. K. contributed equally to this work.

\bibliography{Nonhermitianreviewbib}

\end{document}